\documentclass[smallextended,envcountsect,envcountsame]{amsart} 

\usepackage{graphicx}
\usepackage{enumerate}\usepackage{anysize}
\usepackage{fancyhdr}
\usepackage{amsmath}
\usepackage{amssymb}
\usepackage{amsxtra}
\usepackage{amsfonts}
\usepackage{bbold}
\usepackage{subfigure}
\usepackage{longtable}
\usepackage{caption}
\usepackage{tabularx}
\usepackage[table]{xcolor}
\usepackage{tikz}
\usepackage{comment}

\numberwithin{equation}{section}
\numberwithin{table}{section}
\newtheorem{thm}{Theorem}[section]
\newtheorem{lem}[thm]{Lemma}
\newtheorem{cor}[thm]{Corollary}
\newtheorem{prop}[thm]{Proposition}

\newtheorem{defin}[thm]{Definition}
\newtheorem{rem}[thm]{Remark}
\newtheorem{rems}[thm]{Remarks}

\newcommand{\eee}{{\rm e}}

\newcommand{\one}[1]{\mathbb{1}_{\left\{#1\right\}}}
\newcommand{\card}[1]{\left| #1 \right|}
\newcommand{\bbd}[1]{\boldsymbol{#1}}

\newcommand{\X}{\mathbb{X}}

\newcommand{\hsp}{\mathcal{H}}

\newcommand{\tr}{\mathrm{tr}}
\newcommand{\wtr}{\widehat{\mathrm{tr}}}
\newcommand{\Tr}{\mathrm{Tr}}
\newcommand{\pro}{\mathcal{P}}
\newcommand{\fpro}{\mathcal{F}}
\newcommand{\B}{\mathcal{B}}
\newcommand{\alg}{\mathcal{A}}
\newcommand{\sset}{\subset\!\subset}
\newcommand{\lat}{\mathbb{L}}
\newcommand{\unit}{\mathbf{1}}
\newcommand{\empbf}[1]{{\bf \emph{#1}}}
\newcommand{\supp}{{\rm supp}}

\newcommand{\bonds}{\B_\lat}

\def\tende#1{\vtop{\ialign{##\crcr\rightarrowfill\crcr
              \noalign{\kern-1pt\nointerlineskip}
              \hskip3.pt${\scriptstyle #1}$\hskip3.pt\crcr}}}

\begin{document}
\title[HTE for quantum lattice spin systems]{High-temperature cluster expansion for classical and quantum spin lattice  systems with multi-body interactions}
\author{Nguyen Tong Xuan}
\address{New York University Shanghai, 1555 Century Avenue, 
Pudong New Area, Shanghai, 
China 200122}
\email{tn2137@@nyu.edu}
\author{Roberto Fern\'andez} 
\address{Utrecht University (emeritus) and  New York University Shanghai, 1555 Century Avenue, 
Pudong New Area, Shanghai, 
China 200122}
\email{rf87@nyu.edu}

\date{Received: date / Accepted: date}

\thanks{}
\maketitle
\begin{abstract} 
We develop a novel cluster expansion for finite-spin lattice systems subject to multi-body quantum ---and, in particular, classical--- interactions.  Our approach is based on the use of ``decoupling parameters", advocated by Park \cite{Park82}, which relates partition functions with successive additional interaction terms.  Our treatment, however, leads to an explicit expansion in a $\beta$-dependent effective fugacity that permits an explicit evaluation of free energy and correlation functions at small $\beta$.  To determine its convergence region we adopt a relatively recent cluster summation scheme that replaces the traditional use of  Kikwood-Salzburg-like integral equations by more precise  sums in terms of particular tree-diagrams  \cite{BFP10}.  As an application we show that our lower bound of the radius of $\beta$-analyticity is larger than Park's for quantum systems two-body interactions. 
\end{abstract}

\section{Introduction and overview} 

Cluster-expansion technology has played a central role in the development of rigorous perturbative arguments in statistical mechanics and field theory (see, e.g.\ \cite{BF76,{GJS73}}).  
In its canonical form, a cluster expansion is an expansion, in powers of the fugacity $z$, of the logarithm of the (grand-canonical) partition function of a gas of objects subjected to two-body interactions.  The fugacity is, in fact, an effective parameter whose expression depends on the application.  For instance, in the original setup of real gases $z$ is the exponential of the inverse temperature times the Gibbs chemical potential.  The basic combinatorics, and much of the available theory, derives from the relatively simpler case of interactions corresponding to some form of purely hard-core exclusion.   Cluster-expansion technology amounts, in general, to rewriting a series so it takes the form of a cluster expansion of systems with pure exclusions and some appropriate fugacity. 

The main role of cluster expansions is to determine small-fugacity regions of analyticity of thermodynamical potentials.  Through differentiation, this in turn implies the analyticity of (reduced) correlation functions and, through additions of suitable interaction terms, the analyticity of expectations of more general observables.  For this role, the expansion competes with the Kirkwood-Salzburg (KS) technique based on systems of equations among correlations \cite{galmir68,GK71} (see also \cite[Chapter 4]{Rue69}).  This last approach does not makes explicit use of the expansion ---though its existence is tacitly assumed--- and analyticity of the correlations is obtained by defining an artificial Banach structure and resorting to a fixed-point argument.  A posteriori, however, the analyticity of correlations implies the convergence of the cluster expansion and the analyticity of the corresponding thermodynamical potential.  While direct treatment of  the expansion leads to more explicit results, the use of KS equations yields analyticity results almost equivalent to those obtained in the best inductive proofs of convergence of cluster expansions \cite{BFP10}.  The KS approach, however, can not benefit from newer expansion summation algorithms \cite{PF07,PY17,FN19,fia20} which result in larger analyticity domains. 

In classical statistical mechanics both techniques ---KS equations and cluster expansions--- have been systematically used to study the high-temperature region of phase diagrams.  These studies should be, in principle, based on  the canonical high--temperature expansions (HTE) which is obtained by successive derivatives of the free energy with respect to the inverse temperature $\beta$.  This is, however, almost never the case because  the terms of this expansion are given by truncated expectations, also known as Ursell functions or cumulants (see e.g.~\cite[Section II.12]{Simon93}), which are difficult to bound because they involve sums of terms with alternating signs. The KS approach bypasses these combinatorial issues and is naturally applicable to general summable interactions, regardless of its many-bodynes~\cite{galmir68,isr76}. In contrast, expansion techniques involve a first stage in which the HTE is rewritten in a form more amenable to termwise control involving powers of parameters that are  decreasing functions of $\beta$. These alternative ``high-temperature expansions"  can be summed through a variety of techniques, for instance using random-walk bounds~\cite[Section V.6]{Simon93}.  The expansions leading to more precise results, however, are those written in terms of \emph{polymers} interacting purely through exclusion conditions (see, e.g.~\cite[Section V.7]{Simon93}). 
The application of these polymer expansions has been, however, restricted to two-body interactions~\cite{domgre74}, with the exception of references \cite{prosco00,JT19} that, however, rely on a dominant two-body component.  There is, therefore, the need for a  high-temperature cluster expansion applicable to classical systems with general multi-body interactions.
 
In quantum statistical mechanics, low-fugacity and high-temperature studies differ according to whether the system is in continuum space or in lattices. In the former case all existing references resort to the Feynaman-Kac representation in terms of Wiener integrals.  Initial rigorous low-fugacity work by Ginibre~\cite{gin65.1,gin65.2,gin65.3} (see surveys in \cite[Section 4.6]{Rue69} and \cite{gin72}) make use of Kirkwood-Salzburg equations. In more recent work~\cite{Uel04,PU09}, the same representation is used to obtain high-temperature and high-dilution expansions.  All these references, however, focus on quantum gases subject to stable pair interactions.  For lattice systems most of the references employ the Kirkwood-Salzburg approach.  This includes the original work of Greenberg on high-temperature analyticity~\cite{Green69a,Green69b} and that of Park and Yoo on unbounded boson systems ~\cite{paryoo95}.  The former works for general many-body exponentially summable interactions while the latter ---which uses the Wiener integral representation--- only applies to two body terms.  We also mention the expansion proposed by Kennedy~\cite{Ken85}, based on the Trotter formula, and that in  \cite{DFF96}, based on the Duhamel identity.  These last two expansions, however, are adapted only to low-temperature applications.  

Park's first paper~\cite{Park82} is the only published work referring to a quantum many-body cluster expansion applied to high-temperature studies.  His cluster expansion, however, is written in terms of ratios of partition functions and is not fully developed so to make contact with usual polymer expansions amenable to available summation techniques.   Convergence is proven through bounds involving the Peierls-Bogoliubov inequality, and the analyticity of thermodynamic quantities is proven through KS equations.  The need remains, therefore, for a more efficient full-fledged high-temperature quantum cluster expansion.

The expansion derived in this paper fills the lack of multy-body cluster expansions for both  classical and quantum lattice spin systems.  Our presentation is done in a rather general framework that includes both types of spin models and that clearly specifies which mathematical properties are required.  While statistical mechanics spaces are endowed with a rich mathematical structure, not all of it is needed in the analysis below.  In particular, the topology of classical spin spaces is irrelevant and self-adjointness of the quantum mechanical operators plays only an ancillary role.  This last fact is expected in analyticity studies that involve extending the values of parameters to the complex domain.

Our analyticity results require interactions to be exponentially summable in the number of bodies, which is an optimal requirement given the level of generality~\cite{dobmar88}.  We remark that for classical finite-spin models, there exist high-temperature analyticity results for plainly summable interactions \cite{galmir68,isr76}.  These results, however, correspond to a more restricted notion of analyticity associated to smaller space of interactions~\cite{entfer89}.  
For quantum spin systems, exponential summability is needed already for the existence of an analytic quantum dynamics (see, e.g., Theorems 6.2.4 and 6.2.9 in \cite{brarob2}) and for the equivalence between Gibbs and KMS states (Corollary 6.2.19 in \cite{brarob2}).

Our expansion is inspired in, and uses ideas of Park~\cite{Park82}, themselves related to the scheme proposed by Greenberg~\cite{Green69a}.  The approach is based on 
three mathematical ingredients, the first two obtained by disentangling Greenberg's and Park's proposals:

\begin{itemize}
\item[(i)] \emph{Canonical way to turn a model into a gas with hard-core exclusions.} The model must contain some (fugacity) weights that factorize whenever a given compatibility relation is met.  In Park's approach this rewriting is achieved through the use of decoupling parameters.  This technique is already present in the early field-theoretical cluster expansions, but is infrequently used in classical expansions due to the efficiency of graph-theoretical encodings of the expansion terms.  We show, however, that the underlying mechanism is simply M\"obius transform. Furthermore, as we emphasize in Proposition \ref{prop:gas1} and Theorem \ref{thm:polmect1.1}, it is more natural to obtain the gas expression at the level of operators, so it can subsequently be subjected to different linear functionals to obtain different results.

\item[(ii)] \emph{Reliance on trace factorization.}  The theory is built only on the validity of the factorization of the trace with respect to tensor products.  On the one hand this prevents its application to quantum itinerant particles, because the  factorization is lost upon projection into symmetric (bosons) or antisymmetric (fermions) subspaces.  On the other hand, it makes the theory applicable to classical systems, for which ``trace" corresponds to integration with respect to an a-priori product probability measure.  

\item[(ii)] \emph{Use of well known convergence criteria.}  The expansion allows for the immediate use of cluster-expansion summability criteria, in particular the more recent improved criteria~\cite{PF07,PY17,FN19,fia20}. The domain of analyticity is determined on the basis of the pinned expansion, instead of the Peierls-Bogoliubov inequality used by Park.  
\end{itemize}

The resulting cluster expansion (Theorem \ref{thm:polmect1.1}) is an expansion in polymers defined by collections of bonds, that is, by finite subsets of vertices of a lattice.  The fugacity associated to each collection has a relatively complicated expression [formula \eqref{eq:ms2}] which is, however, natural in light of the M\"obius transform.  Despite its apparent complexity, the expansion yields full control ---and explicit series expressions--- for the free-energy density and observable expectations.  Moreover, it yields the absence of phase transitions in the strongest possible sense, namely implying both uniqueness of (no necessarily translation-invariant) Gibbs states and analyticity of free energy and expectations.   

The convergence and analyticity results obtained through our approach are presented below with different levels of generality.  The most general and precise version is presented in Theorem~\ref{theo:rr.tan.06}, which should be considered our main analyticity result.  A less abstract, but weaker version is given in Theorem~\ref{corollary1.1}.  The most concrete, and weakest, version is the content of Proposition \ref{prop.xuan.1}, which provides an explicit criterion in terms of interaction terms.  Issues involving boundary conditions ---in particular the uniqueness result--- are only stated in this more concrete set up.
We emphasize that neither of these results require the interaction to be translation invariant.
For completeness, the paper also presents a final section describing an improved Kirkwood-Salzburg approach which, resorting to the alternate fixed-point argument presented in~\cite{BFP10}, almost reaches the criterion of  Corollary~\ref{corollary1.1}.   

\section{Setup}

\subsection{{\bf Spins, observables and states}}
We consider a countable set of points (sites) called the \emph{lattice} $\lat$.  At each site $x\in\mathbb L$ sits a space $\hsp_x$ whose structure depends on the application, as discussed below. We shall call $\hsp_x$ the \emph{spin-space} at $x$.   In most cases all these spaces are copies of a fixed  space $\hsp$ and $\lat$ is such that  $\mathbb{Z}^d$ acts injectively or bijectively, defining the notions of periodicity or translation invariance (in particular, $\mathbb L$ may coincide with  $\mathbb{Z}^d$ itself).  This is not required in most of the following (the exception is the definition of the free energy density, Section \ref{ssec:r.free-energy}).
We shall denote $\fpro$ the set of non-empty finite subsets of $\mathbb L$ and $\fpro(\Lambda)$  
the family of non-empty subsets of $\Lambda\in\fpro$.   
For each $\Lambda\sset \mathbb L$ we denote 
\begin{equation}
\hsp_{\Lambda} \;=\; \otimes_{x\in\Lambda}\hsp_{x}
\end{equation}
so that $\hsp_{\Lambda_1\cup\Lambda_2} = \hsp_{\Lambda_1}\otimes\hsp_{\Lambda_2}$ if $\Lambda_1\cap\Lambda_2=\emptyset$ .
The main ingredients of the formalism are the following.

\begin{description}
\item[\empbf{Algebras of observables}]  A sequence of complex unital Banach algebras $\{\alg_\Lambda: \Lambda\in\fpro\} $ of functions on $\hsp_\Lambda$.  The algebras are also required to be $C^*$-algebras, but the involution only plays a role in the definition of state below.  In particular, the notion of self-adjointness is not invoked inn the sequel.  The sequence $\{\alg_\Lambda: \Lambda\in\fpro\} $ is required to satisfy the following:
\begin{itemize}
\item[(A1)] \emph{Monotonicity:}
  The sequence is increasing with respect to the partial order of $\fpro$ in the sense that
\begin{equation}
\Lambda_1\subset\Lambda_2\in\fpro \;\Longrightarrow\;\alg_{\Lambda_1}\subset\alg_{\Lambda_2}\;
\end{equation}
where the right-hand side means that $\alg_{\Lambda_1}$ is isomorphic to a sub-algebra of $\alg_{\Lambda_2}$.  Explicitly, the isomorphism is defined by the map
\begin{equation}
\alg_{\Lambda_1} \ni A_1\;\longmapsto\; A_1\otimes \unit_{\Lambda_2\setminus\Lambda_1}\in\alg_{\Lambda_2}\;
\end{equation}
where $\unit_\Lambda$ indicates the unit of $\alg_\Lambda$.
\item[(A2)] \emph{Disjoint commutativity:} If $A_1\in \alg_{\Lambda_1}$ and $A_2\in \alg_{\Lambda_2}$, with $\Lambda_1,\Lambda_2\subset\Lambda$,
\begin{equation}
\label{ea:rfr.1}
\Lambda_1 \cap \Lambda_2=\emptyset \; \Longrightarrow\; [A_1,A_2]=0\;.
\end{equation}
\end{itemize}

The monotonicity property implies that the 
\begin{equation}
\alg^{\mathrm loc}\;=\; \bigcup_{\Lambda\in\fpro} \alg_\Lambda
\end{equation}
is well defined as a projective limit of the $\alg_\Lambda$ and constitutes an algebra called the \empbf{algebra of local observables}.  It is, however, not complete; its norm completion
\begin{equation}
\alg\;=\; \overline{\bigcup_{\Lambda\in\fpro} \alg_\Lambda}^{\|\;\|}
\end{equation}
is the \empbf{algebra of quasilocal obesrvables}.  This algebra embodies the system attained as $\Lambda\to\lat$ (thermodynamic limit).
\smallskip

The \empbf{support} of a local observable $A$ is the minimum, in the set-inclusion order, of the sets $\Lambda\in\fpro$ such that $A\in\alg_\Lambda$.  Alternatively, it is the intersection of all the finite $\Lambda$ with this last property.  It will be denoted $\supp(A)$.
\medskip

\item[\empbf{Trace operations}] A sequence of linear forms $\tr_\Lambda: \alg_\Lambda\longrightarrow \mathbb{C}$, satisfying the following:

\begin{itemize}
\item[(T0)] \emph{Tracial property:} If $A, B\in\alg_\Lambda$
\begin{equation}
\tr_\Lambda(AB)=\tr_\Lambda(BA)\;.
\end{equation}

\item[(T1)] \emph{Normalization:} 
\begin{equation}
\tr_\Lambda(\unit_\Lambda)=1\;.
\end{equation}

\item[(T2)] \emph{Continuity:} If $A\in\alg_\Lambda$
\begin{equation}
\left|\tr_\Lambda(A)\right|\;\le\;\|A\|\;.
\end{equation}

\item[(T3)] \emph{Factorization:} For disjoint $\Lambda_1,\Lambda_2\in\fpro$ and  $A_1\in\alg_{\Lambda_1}$, $A_2\in\alg_{\Lambda_2}$
\begin{equation}\label{eq:rfr.2}
\tr_{\Lambda_1\cup\Lambda_2}\bigl(A_1\otimes A_2\bigr)\;=\; \tr_{\Lambda_1}\bigl(A_1\bigr)\, \tr_{\Lambda_2}\bigl(A_2\bigr)\;.
\end{equation}
This property is, in fact, equivalent to 
\begin{equation}
\tr_\Lambda\;=\; \bigotimes_{x\in\Lambda} \tr_{\{x\}}\;.
\end{equation}
\end{itemize}

Normalization plus factorization implies that $\tr_{\Lambda}(A)$ is independent of $\Lambda$ as long as $A\in\alg_{\Lambda}$.  Hence, the factorization \eqref{eq:rfr.2} yields that, if  $\Lambda_1, \Lambda_2\subset\Lambda$, with $\Lambda_1\cap\Lambda_2=\emptyset$, then, for all operators $A_1$ on $\hsp_{\Lambda_1}$ and $A_2$ on $\hsp_{\Lambda_2}$,
\begin{equation}\label{eq:rr.sl.0}
\tr_\Lambda(A_1A_2)\;=\; \tr_\Lambda(A_1)\, \tr_\Lambda(A_2)\;.
\end{equation}

\item[\empbf{Partial traces}] More generally, to consider boundary conditions, traces must allow ``tracing out" partial spin configurations.  Formally, we assume that the previous maps $\tr_\Lambda: \alg_\Lambda\longrightarrow \mathbb{C}$ extends to maps
\begin{equation}
\wtr_\Lambda: \bigcup_{\Gamma\in\fpro}\alg_\Gamma\;\longrightarrow\; \bigcup_{\Gamma\in\fpro_{\lat\setminus\Gamma} }\alg_\Gamma
\end{equation}
such that 
\begin{equation}
\wtr_\Lambda\bigl(A\otimes B\bigr)\;=\; \tr_\Lambda(A)\,B
\end{equation}
for all $A\in\alg_\Lambda$ and $B\in \alg_\Gamma$ with $\Gamma\sset\lat\setminus\Lambda$.  In fact, this property uniquely defines $\wtr_\Lambda$.
\medskip

\item[\empbf{States}]
A \empbf{state} on a $C^*$-algebras is a (possibly complex valued) linear funcional $\Theta$ that is
\begin{itemize}
\item[(T1)] \emph{Normalized:} $\Theta(\unit)=1$.
\item[(T2)] \emph{Positive:} $\Theta(A A^*)\ge 0$ for all operators $A$ in the algebra.
\end{itemize}
It is not difficult to see that these properties imply \cite[Section I.7]{Simon93}
\begin{equation}
\card{\Theta(A)}\;\le\; \|A\|\;.
\end{equation}
States on the algebras $\alg_\Lambda$ ---built on the finite-dimensional spaces $\hsp_\Lambda$--- are of the form $\Theta_\Lambda(A)=\tr\bigl(\theta_\Lambda A\bigr)$
for some operator $\theta_\Lambda\in\alg_\Lambda$ that is also positive and normalized.  States on the quasilocal algebra $\alg$ are uniquely defined, by density,  by their restrictions to the different local algebras $\alg_\Lambda$.  Therefore, they  are uniquely defined by the family of operators $\{\theta_\Lambda : \Lambda\in\fpro\}$.  These operators must, however, satisfy the consistency conditions 
\begin{equation}
\tr_\Lambda\bigl (\rho_\Lambda A\bigr) \;=\; \tr_{\Lambda'}\bigl (\rho_{\Lambda'} A\bigr) \qquad\mbox{if} \quad
A\in\alg_\Lambda \mbox{ and } \Lambda\subset\Lambda'
\end{equation}
or, equivalently,
\begin{equation}
\wtr_{\Lambda'\setminus\Lambda} \bigl(\rho_{\Lambda'}\bigr)\;=\; \rho_\Lambda \qquad \mbox{if} \quad \Lambda\subset\Lambda'\;.
\end{equation}
Conversely, for every family of normalized positive local operators $\theta_\Lambda\in\alg_\Lambda$ satisfying the previous consistency conditions,
there exists a unique state $\Theta$ in $\alg$ whose restrictions to each $\Lambda$ are defined by the corresponding $\theta_\Lambda$.

\medskip

Let us now precise the previous construction for the two cases of interest for us.

\end{description}

\begin{description}
\item[\bf Classical spin systems]   Each $\hsp_x$ is a measure space with an a-priori probability measure $d\mu_x$.  The algebra $\alg_\Lambda$ is the set of bounded measurable complex-valued functions on $\hsp_\Lambda$.  The trace operation is simply the integration with respect to the product measure: $\tr_\Lambda=\otimes_{x\in\Lambda}d\mu_x=\wtr_\Lambda$.  Two comments are in order.  First, the true statistical mechanical observables are real-valued functions, but complex values must be included to discuss analyticity.  Second, it is customary to demand in each $\hsp_x$ a topology compatible with the measure structure (that is, such that continuous functions are measurable) and, furthermore, making each $\hsp_x$ a compact space. Such topological properties play no role in the sequel, so we ignore them.  An advantage of the classical formalism is that there is a well-defined notion of full configuration space, namely $\otimes_{x\in\lat} \hsp_x$, and quasilocal observables are continuous functions in this infinite-volume space.  Classical statistical mechanics notions, however, are defined through finite-volume approximations or through conditions involving objects defined in finite volumes (DLR equations).   Classical states are defined by probability measures.  For finite regions they are absolutely continuous with respect to the product measure, but this is no longer the case for the whole of $\lat$.  

\item[\bf Quantum spin systems]  $\hsp$ is a finite-dimensional Hilbert space and $\alg_\Lambda$ is the set of bounded operators on $\hsp_\Lambda$.  The trace operation is just the normalization
\begin{equation}\label{eq:rr-trace}
\tr_{\Lambda}\;=\;\frac{1}{\dim\hsp_{\Lambda}}\Tr_{\hsp_{\Lambda}}
\end{equation}
of the canonical trace $\Tr_{\hsp_{\Lambda}}$ of $\hsp_\Lambda$.
The associated partial traces are more easily defined by choosing an orthonormal basis $\{e_i: i\in I_\Lambda\}$ of $\hsp_\Lambda$ and writing each operator $C\in\alg_{\Lambda\otimes \Gamma}$ in the form $C=\{C_{ij}: i,j\in I_\Lambda\}$ with $C_{ij}\in\alg_{\Gamma}$.  With this,
\begin{equation}\label{eq:rfr.3}
\wtr_\Lambda(C)\;=\; \frac{1}{\card{I_\Lambda}} \sum_{i\in I_\Lambda} C_{ii}\;.
\end{equation}
The isometric isomorphism between $ \hsp_\Lambda\otimes \hsp_\Gamma$ and $\bigoplus_{i\in I_\Lambda} \bigl(\mathbb C e_i \otimes \hsp_\Gamma\bigr)$ implies that \eqref{eq:rfr.3} is independent of the basis chosen for $\hsp_\Lambda$. 
Unlike the classical case, there is not a well defined infinite-volume ``configuration space" because a countable product of Hilbert spaces is not univocally defined.  For this reason, the relevant infinite-volume objects are the quasilocal observables themselves, defined by projective limits.   
As mentioned above, quantum states on finite regions are given by traces weighted by local operators $\theta_\Lambda$.  States on the whole of $\lat$ can not, in general, be written in terms of a single global operator weight.

\end{description}

\subsection{{\bf Interactions and Hamiltonians}}
An interaction is a family $\Phi=\{\Phi(X):X\in\fpro\}$  with $\Phi(X)\in\alg_X$.  Most operators in an interaction are zero except for a limited family.  It is traditional to call \empbf{bonds} the supports of nonzero interaction terms.  We denote
\begin{equation}
\bonds\;=\;\; \bigl\{ X\in\fpro : \Phi(X) \neq 0\bigr\} \;.
\end{equation}

Our treatment requires interactions to satisfy the following summability condition.
\begin{equation}\label{eq:intro1}
\exists\, \alpha>0 \quad:\quad \|\Phi\|_\alpha\;:=\;
\sup_{x\in \mathbb{Z}^d}\sum_{\substack{X\in\fpro \\ x\in X}}\left\|\Phi(X)\right\|\eee^{\alpha|X|}<\infty
\end{equation}
for some $\alpha>0$. 
This is precisely the condition invoked by Park in his KS treatment.  As commented above, for classical interactions Dobrushin and Martirosyan~~\cite{dobmar88} showed that, at the present level of generality, such a condition is not only sufficient but also necessary for the analyticity of the free energy and observables.    In the quantum case, condition \eqref{eq:intro1} is necessary for the existence of the dynamics on which the definition of KMS states relies.  We shall denote $\mathbb B^{(\alpha)}$ the Banach space of interactions with finite $ \|\Phi\|_\alpha$.   

Each interaction defines  Hamiltonians
\begin{equation}\label{eq:rr.qq.1}
H_\Lambda\;=\; \sum_{X\in\B_\Lambda} \Phi(X)\;,
\end{equation}
corresponding to free boundary conditions and, more generally, 
 extended hamiltonians
\begin{equation}\label{eq:rr.qq.1.1}
\overline H_\Lambda\;=\; \sum_{X: X\cap\Lambda\neq\emptyset}\Phi(X)\;,
\end{equation}
to be used to prove uniqueness of Gibbs states.  Condition \eqref{eq:intro1} implies that the possibly infinite sum in the right-hand side does exist (in fact, already for $\alpha=0$).  To be more precise, let us introduce partial Hamiltonians
\begin{equation}
H_{\B}\;=\; \sum_{X\in\B} \Phi(X)
\end{equation}
for every family $\B\subset\bonds$.  In particular, 
for each set $\Lambda\subset\lat$ we denote
$\B_\Lambda=\bonds\cap\Lambda=\{X\subset\Lambda\,|\, \Phi(X)\neq 0\}$ and
\begin{equation}
\B_{\partial\Lambda} \;=\; \bigl\{ X\in \fpro \bigm| X\cap\Lambda\neq\emptyset\;,\;X\cap\lat\setminus \Lambda\neq\emptyset\;,\; \Phi(X)\neq 0\big\}\;.
\end{equation}
Hence, $H_\Lambda\equiv H_{\B_\Lambda}$ and we have the decomposition
\begin{equation}\label{eq:rr.qq.1.4.0}
\overline H_\Lambda\;=\; H_\Lambda + W_\Lambda\qquad\mbox{with}\quad 
W_\Lambda\;=\; \sum_{X\in\B_{\partial\Lambda}}  \Phi(X)\;.
\end{equation}
The termwise norm of the last sum is bounded above by $\card{\Lambda}\|\Phi\|_0\,$, ensuring convergence.

More generally, it is necessary to consider \empbf{(quantum) boundary conditions} defined by states $\Theta$ on $\alg_{\lat\setminus\Lambda}$.  
%
In the classical case, the probabilistic machinery allows to consider only ``frozen spin" conditions which correspond to projections on a single element of an orthonormal basis or, more concretely, to delta functions of the form $\delta{\sigma_X}$ for some spin configuration $\sigma$ on $\lat\setminus\Lambda$.  General states must be considered, however, for the quantum case.

The boundary conditions yield ``conditioned" Hamiltonians 
\begin{equation}\label{eq:rr.qq.1.4.1}
 H_\Lambda^\Theta \;=\; H_\Lambda + W_\Lambda^\Theta 
\end{equation}
where the boundary term is obtained by ``$\Theta$-tracing out" the boundary spins:
\begin{eqnarray}\label{eq:rr.qq.1.4.2}
W_\Lambda^\Theta &=& \sum_{X\in\B_{\partial\Lambda}}  \wtr_{X\setminus\Lambda}\bigl[\bigl(\one{X\cap\Lambda}\otimes \theta_{X\setminus\Lambda}\bigr)\,\Phi(X)\bigr] \nonumber\\
&=:&   \sum_{X\in\B_{\partial\Lambda}} \Phi^\Theta(X)\;.
\end{eqnarray}
Notice that this series converges because it is absolutely and uniformly bounded by $\card{\Lambda}\|\Phi\|_0$.

In the case $\lat=\mathbb Z^d$, \empbf{periodic boundary conditions} are useful for arguments requiring translation invariance and for numerical simulations.  They are defined for rectangular regions $\Lambda_{\underline a}$ with sides of sizes $\underline a=(a_1,\ldots,a_d)$.  Informally they are obtained by letting boundary bounds to ``wrap around" by periodicity.  Formally, one considers the map $\tau_{\underline a}: \mathbb Z^d \to \Lambda_{\underline a}$ such that $\tau_{\underline a}(x)$ is the unique site in $\Lambda_{\underline a}$ equal to $x$ module $\underline a$.  This map, in turns, defines a map from finite sets in $\mathbb Z^d$ to subsets of $\Lambda_{\underline a}$ with
$\tau_{\underline a}(X)=\cup_{x\in X} \tau_{\underline a}(x)$
(we use the same notation as there is no risk of confusion) 

The periodic boundary conditions are carried by the bonds in
\begin{equation}
\B^{\mathrm per}_{\partial_{\underline a}}\;=\; \Bigl\{  \mbox{single }X\in\B_{\partial\Lambda} : 
 \tau_{\underline a}\bigr|_X \mbox{ is one-to-one} \Bigr\} \;.
\end{equation}
The word ``single" encodes the fact that, for translation-invariant interactions, each set $X\in\B_{\partial\Lambda}$ has another copy, differing by a translation by $\pm\underline a$, with the same wrap-around restriction.  In this case, only one of these copies has to be considered. 
The requirement of one-to-one-ness prevents bonds whose periodic wrapping covers some site twice or more times.  This is usually not a restriction, if $\underline a$ is sufficiently large, for finite-range interactions ---which are the ones involved in usual simulations.  The condition is, however, needed in general to have well defined maps $ [\tau_{\underline a}|_X]^{-1}$ from $\hsp_{\tau_{\underline a}(X)}$ to $\hsp_X$ mapping ---for an arbitrary choice of orthonormal basis--- the vector with components 
$\omega_y:y\in \tau_{\underline a}(X)$ to the vector with components $\widetilde \omega_x= \omega_{\tau_{\underline a}(x)}:x\in X$.  

The periodic Hamiltonian is, then,
\begin{equation}\label{eq:rr.qq.1.4}
 H_\Lambda^{\mathrm per}\;=\; H_\Lambda + W_\Lambda^{\mathrm per}
 \end{equation}
 with
 \begin{equation}
W_\Lambda^{\mathrm per}\;=\; \sum_{X\in \B^{\mathrm per}_{\partial_{\underline a}}}  \Phi(X)\circ [\tau_{\underline a}|_X]^{-1} \;=\; \sum_{X\in \B^{\mathrm per}_{\partial_{\underline a}}}  \Phi^{\rm per}(X) \;.
\end{equation}
On a formal level, $H_\Lambda$ corresponds to the particular case of \emph{free boundary conditions}, and is sometimes denoted as $H^\emptyset_\Lambda$.  Conversely, both $H^\Theta_\Lambda$ and $H^{\mathrm per}_\Lambda$ can be considered Hamiltonians in $\Lambda$ with free boundary conditions, with interactions increased by the projection on $\Lambda$ of the $\Theta$-traced out or the wrapped-around boundary terms. 

\subsection{{\bf Finite-volume Gibbs states and free energies}}
The \empbf{The Gibbs state on $\hsp_\Lambda$ with free boundary conditions} is the linear form on $\alg_\Lambda$ defined by 
\begin{equation}\label{eq:rr.qq.2}
\pi_\Lambda(A)\;=\; \frac{\tr_\Lambda \bigl(A \,\eee^{-\beta H_\Lambda}\bigr)}{Z_\Lambda}
\end{equation}
for any $A\in\alg_\Lambda$.  The normalization
\begin{equation}\label{eq:rr.qq.3}
Z_\Lambda\;=\; \tr_\Lambda \bigl(\eee^{-\beta H_\Lambda}\bigr)
\end{equation}
is the \emph{partition function} and $\beta$ is the inverse temperature.  In the classical case \eqref{eq:rr.qq.2} is, in fact, an expectation.
Analogously, the \empbf{Gibbs state on $\hsp_\Lambda$ with  boundary conditions $\Theta$} is the linear form on $\alg_\Lambda$ defined by 
\begin{equation}\label{eq:rr.qq.4}
\pi_\Lambda^\Theta(\bullet )\;=\; \frac{\tr_\Lambda \bigl(\bullet  \,\eee^{-\beta H_\Lambda^\Theta}\bigr)}{Z_\Lambda^\Theta} \qquad \mbox{with}\quad 
Z_\Lambda^\Theta\;=\; \tr_\Lambda \bigl(\eee^{-\beta H_\Lambda^\Theta}\bigr)\;,
\end{equation}
and the \empbf{Gibbs state on $\hsp_\Lambda$ with periodic boundary conditions} is defined by the linear form on $\alg_\Lambda$ 
\begin{equation}\label{eq:rr.qq.5}
\pi^{{\mathrm per}}_\Lambda(\bullet)\;=\; \frac{\tr_\Lambda \bigl(\bullet  \,\eee^{-\beta H_\Lambda^{\mathrm per}}\bigr)}{Z^{\mathrm per}_\Lambda} \qquad \mbox{with}\quad 
Z^{{\mathrm per}}_\Lambda\;=\; \tr_\Lambda \bigl(\eee^{-\beta H_\Lambda^{\mathrm per}}\bigr)\;,
\end{equation} 
\medskip

Cluster expansions are developments of the \emph{finite-volume free energies}
\begin{equation}
-\beta F^\#_\Lambda\;=\ \log Z^\#_\Lambda
\end{equation}
as a formal power series of suitable effective fugacities.  Here $\#=\emptyset,\Theta,{\rm per}$. 
The expansions are high-temperature if these fugacities tend to zero as $\beta\to 0$.  Analyticity of the infinite-volume expectations and of the free-energy density follows by proving that the series converges absolutely within a fugacity polydisc whose radii are uniform in $\Lambda$.  
The proof of the existence and analyticity of the limit free-energy densities and Gibbs states is, therefore, the same in all cases, but the contribution of the boundary terms must be accounted for to show that these limits are independent of the boundary conditions.  In fact, general arguments show that free energy densities always exist and are independent of the sequence of exhausting volumes ---as long as they form a van Hove sequence--- and of the boundary conditions (see, e.g., \cite[Sections 1.2--1.3]{isr76} and \cite[Sections II.2--II.3]{Simon93}). Cluster expansions are useful to determine regions of analyticity and to provide explicit expressions ---in the form of power series--- of free energies and Gibbs expectations.
\medskip

\subsection{{\bf (Infinite-volume) Gibbs  = KMS states}}\label{sec:gibbs-KMS}

In the sequel, unless otherwise stated,  the infinite-volume limits are understood in the sense of set inclusion.  
\begin{defin}[Convergence in the sense of inclusion]
A family of objects $F_\Lambda$, belonging to a Banach space (e.g.\ complex numbers, finite dimensional operators) and labelled by a family of finite sets $\Lambda\in\mathcal S$--- \empbf{converges in the sense of set inclusion} to a target object $F$ if
\begin{equation}
\forall \varepsilon \; \exists \Lambda_\varepsilon\in\mathcal{S} \mbox{ such that } \|F_\Lambda-F\|<\varepsilon\;\forall \Lambda\supset\Lambda_\varepsilon\,,\; \Lambda\in\mathcal{S}\;.
\end{equation}
\end{defin}
The default situation is when $\mathcal{S}=\fpro$, the family of all the finite subsets of the lattice.
For the convergence of the free-energy density (Section \ref{ssec:r.free-energy}), limits will involve the most restrictive van Hove convergence, which only holds for suitable families of finite subsets of the lattice.  
\smallskip
 
In the classical case, the theory of Gibbs states benefits from a sophisticated probabilistic machinery that leads to two ways of characterizing Gibbs states on the full lattice: (1) Through the DLR equations, and (2) through limits $\Lambda\to\lat$ of the states $\pi^\beta_\Lambda(\bullet \mid \Theta)$.  These limits, in particular, realize the extremal points of the simplex of Gibbs states, which are the states representing macroscopic phases.  For the quantum case, the theory of operator algebras takes the place of probability and the situation becomes notoriously more complicated.  Indeed, static DLR-like equations are out of the question, basically because ---as remarked in \cite[Section IV.4]{Simon93}--- $\exp(-\overline H_\Lambda) \neq
\exp(- H_\Lambda)\, \exp(- W_\Lambda)$.  Instead, these equations are replaced by  \emph{KMS conditions} which are dynamical commutation conditions (involving complex values of $\beta$) which amounts to fluctuation-dissipation conditions embodying a sense of local thermodynamic stability \cite[Section 3.7]{Sewell86}.  Unlike DLR equations, KMS conditions apply to objects already at infinite volume, without any reference to finite-volume kernels.  Furthermore, while it is true that the limits of states \eqref{eq:rr.qq.2}, \eqref{eq:rr.qq.4} and \eqref{eq:rr.qq.5}, are indeed KMS states, at the present level of knowledge there is no guarantee that they include the extremal states needed to describe macroscopic phases.  


\begin{defin}
A state $\Theta$ on $\alg$ is a \empbf{Gibbs state} for an interaction $\Phi\in\mathbb B^{(\alpha)}$, for some $\alpha>0$, if for each finite $\Lambda$ there exists a state $\widetilde\Theta_{\lat\setminus\Lambda}$ on $\alg_{\lat\setminus\Lambda}$ such that
\begin{equation}
\Theta(A) \;=\; \lim_{\Lambda'\to\lat} \frac{\pi_\Lambda \otimes \widetilde\Theta_{\lat\setminus\Lambda}\bigl(A\,\eee^{-\beta H_{\Lambda'}}\,\eee^{\beta (H_{\Lambda'}-W_{\Lambda,\Lambda'})}\bigr)}
{\pi_\Lambda \otimes \widetilde\Theta_{\lat\setminus\Lambda}\bigl(\eee^{-\beta H_{\Lambda'}}\,\eee^{\beta (H_{\Lambda'}-W_{\Lambda,\Lambda'})}\bigr)}
\end{equation}
\end{defin}

This definition is equivalent to, but looks different from, those presented in the literature (\cite{araion74}, \cite[Sections 5.3--5.4]{brarob2}, \cite[Sections IV.4--IV.5]{Simon93}).  In particular, it is equivalent to the definition of KMS states (see \cite{araion74}, \cite[Sections 5.4]{brarob2} and \cite[Sections IV.5]{Simon93}).


\section{Results I: Cluster expansions}\label{ss:rr.qmb}

Cluster expansion approaches involve two distinct steps.  
\begin{itemize}
\item[(C1)] \emph{Gas expansion.}  The relevant mathematical quantity ---typically a partition function--- is written as the grand-partition function of a gas of objects subject only to a hard core interaction.  Both the objects and the exclusion hard-core rule can be of a very general nature.  The abstract notion of objects was initially introduced by Gruber and Kunz \cite{GK71} who called them \emph{polymers}.  Dobrushin \cite{Dob96} proposed to call them \emph{animals}, but we shall respect Gruber and Kunz' original nomenclature.  General hard-core interactions were introduced by Koteck\'y and Presiss \cite{KP86} which are, since then, interpreted as \emph{compatibility relations}.  The weights associated to the different polymers are generically called \emph{fugacities}.
\item[(C2)] \emph{Cluster expansion.}  It corresponds to the formal series obtained by taking the log of the gas expansion.  The associated combinatorics is well known and it is, by now, completely standard.  The convergence of the expansion yields a number of physical results, such as lack of phase transitions in its strongest form (analyticity of he free energy) plus complete control ---in the form of perturbative expansions---- of thermodynamical potentials and expectations.  There are, at present, a number of convergence criteria of varying strength and computational convenience.  See \cite{PF07} for a rather general ranking.
\end{itemize}

The essential step in applications is the development of the gas expansion, that is the definition of suitable notions of polymer and compatibility relation.  Once this is done, the standard cluster-expansion machinery takes over and results follow at once.  That is why many authors only discuss the gas expansion and then immediately conclude with the phrase ``and now we apply the cluster expansion", full stop.  Paraphrasing Dobrushin, this amounts to using the cluster expansion as a magical spell.  

\subsection{{\bf General gas expansion}}

The most popular approach to generate a gas expansion in classical statistical mechanics is to use the ``$\pm 1$ trick" which leads to the following expansion:
\begin{eqnarray}\label{eq:can.1}
\exp\Bigl[ -\beta \sum_{X\in \B_\Lambda} \Phi(X)\Bigr] &=& \prod_{X\in \B_\Lambda} \Bigl[1+\bigl(\eee^{-\beta \Phi(X)} -1\bigr)\Bigr]\\
&=& \sum_{\B\subset\B_\Lambda} \xi_\B  \label{eq:can.2}
\end{eqnarray}
with 
\begin{equation}\label{eq:ms2}
    \xi_\B \;=\; \sum_{\widetilde\B\subset\B} (-1)^{\card{\B\setminus\widetilde\B}} \,\eee^{-\beta H_{\widetilde \B}}\;,
\end{equation}
followed by the factorization of $\xi_\B$ according to the maximally connected disjoint components of $\B$.  Our approach relies on the fact that for the quantum case identity \eqref{eq:can.2} remains valid, even when \eqref{eq:can.1} is false.  Hence, the universal underlying ``trick" is, rather, the M\"obius transform a.k.a.\ inclusion-exclusion principle which we recall for completeness. 

\begin{prop}\label{prop:gas1}
Let $\Lambda$ be a finite and $F, G$ functions on the subsets of $\Lambda$ with values on some vector space $\mathbb V$.  Then,
\begin{equation} \label{eq:ms.moeb}
F(C)\;=\; \sum_{B\subset C} G(B) \;, \forall C\subset\Lambda\quad \Longleftrightarrow\quad G(B)=\sum_{A\subset B} (-1)^{\card{B\setminus A}} F(A) \;, \forall B\subset\Lambda
\end{equation}
\end{prop} 
The most direct proof of this equivalence follows, after reversing the order of summation, from the elementary fact that  $\sum_{C\subset D}(-1)^{\card{D\setminus C}}=0$ if $D\neq\emptyset$. this last fact is proven by selecting some $x\in D$ and splitting the sum into two terms, one on the subsets containing $x$ and the other on the subsets not containing it.  It is simple to see that both contributions cancel each other.

The use of the M\"obius identity is somehow hidden in Park's development of Kirkwood-Salzburg equations and, in particular, in the use of decoupling parameters used in pioneer work in quantum field theory.  For completeness, the relation with these techniques is discussed In Appendix \ref{app:decoupling}. 

In its more general form, a gas expansion involves a finite set $\mathcal S$ endowed with a compatibility relation ``$\sim$" between it subsets, satisfying the following properties.  

\begin{defin} Given a set $\mathcal S$, a family of pairs $(A,B)^\sim=:A\sim B\subset \mathcal S^2$  is a \empbf{hard-core compatibility relation} if the following properties hold:    
\begin{itemize}
\item[(A1)] $A\sim B \,\Longleftrightarrow\, B\sim A$,
\item[(A2)] $A\sim B\, \Longrightarrow\, A\cap B=\emptyset$.
\item[(A3)] $A\not\sim A$,
\end{itemize}
\end{defin}
Property (A2) embodies the hard-core condition.  In general, disjointness is not the only condition (that is, the implication is far from an equivalence), otherwise the only connected sets would be the singletons.  Usually, the elements of the $\mathcal S$ have an underlying geometric structure that determines compatibility.  In addition, compatibility may involve further attributes attached to the sets (e.g. colors, spin configurations, etc).  
\begin{defin} Consider a set $\mathcal S$ endowed with a hard-core compatibility relation  ``$\sim$".
\begin{itemize}
\item[(C1)] Two subsets $A$ and $B$ of $\mathcal S$ are said \empbf{compatible} if $A\sim B$, otherwise they are termed \empbf{incompatible} and denoted $A\not\sim B$. 
\item[(C2)] A set $A\subset\mathcal S$ is \empbf{connected} (for the \emph{in}compatibility relation) if it is non-empty and can not be partitioned into two mutually compatible subsets.
\item[(C3)] Finite connected sets are called \empbf{polymers}.  The set of polymers will be denoted $\pro$.
\end{itemize}
\end{defin}

\begin{prop}[General gas expansion]\label{pro:gengas}
Consider a finite set $\mathcal S$ endowed with a hard-core compatibility relation ``$\sim$", an algebra $\mathbb A$ and a function  $F:\{\mbox{parts of } \mathcal S\}\longrightarrow \mathbb A$  such that $F(\emptyset) =1$ and
\begin{equation}\label{eq:fattr}
A \sim B \;\Longrightarrow\; F(A\cup B)=F(A)F(B)=F(B)F(A)\;.
\end{equation}
Then,
\begin{equation}\label{eq:gengas}
F({\mathcal S}) \;=\; \unit + \sum_{n\ge 1} \frac{1}{n!}\sum_{\substack{(B_1,\ldots,B_n)\\ B_i\in\pro\,,B_i \subset \mathcal S }}
\prod_{i=1}^n E(B_i) \prod_{1\le i\le j\le n}  \one{B_i\sim B_j}
\end{equation}
with
\begin{equation}\label{eq:gengas.1}
E(B)\;=\; \sum_{A\subset B} (-1)^{\card{B\setminus A}} F(A)\;.
\end{equation}
\end{prop}
The weights $E(B_i)$ is as the \emph{fugacities} of their respective polymers.  

The proof of \eqref{eq:gengas}--\eqref{eq:gengas.1} is straightforward:   The M\"obius transform \eqref{eq:ms.moeb} leads to
$F({\mathcal S}) = \sum_{B\subset \mathcal S} E(B)$.  The compatibiliy attributes (A1)--(A2) imply that the factorization property \eqref{eq:fattr} is also valid for the fugacities $E(B)$.  Expansion \eqref{eq:gengas} follows, then, by decomposing each $B$ into its maximally connected components.  The sum in the RHS is, in fact, over families $\{B_1,\ldots, B_n\}$, but  
$\sum_{(B_1,\ldots,B_n)}$ can be replaced by $(1/n!) \sum_{\{B_1,\ldots,B_n\}}$ due to the attribute (A3) of the exclusion rule ($B\not\sim B$).
\qed

We notice that, in the above setup, the sum in \eqref{eq:gengas} involves finitely many terms.  Extensions require summability considerations as noted below.    

\subsection{{\bf Operator-valued gas expansions for Boltzmann weights}}

In all our applications, polymers are families of bonds and the compatibility relation between two families is simply disjointness of the unions of their elements.  
\begin{defin} 
\begin{itemize}
\item[(i)] The \empbf{support} of a family $\B=\{X_i\}_i$ of finite subsets of $\Lambda$ is the set
$\underline\B=\cup_i X_i$.
\item[(ii)] Two finite families $\B_1,\B_2$ of finite subsets of $\lat$ are \emph{compatible} if their supports are disjoints ---$\underline \B_1 \cap \underline \B_2 = \emptyset$.
\item[(iii)] A family $\B$ is said \emph{connected} if it is not the union of two mutually compatible families.  
\end{itemize}
\end{defin}

\begin{defin}
The \empbf{polymers} corresponding to an interaction $\Phi$ are finite connected families of bonds $\B$.  The sets of $\Phi$-polymers will be denoted $\pro^\Phi$.  Likewise, the polymers contained in a set $\Lambda\subset\lat$ will be denoted $\pro^\Phi_\Lambda$.
\end{defin}
\smallskip

The most important gas expansion for our purposes is that of the operators $\eee^{-\beta H_\Lambda}$.  They are a straightforward corollary of the general gas expansion of Proposition \eqref{pro:gengas}.

\begin{prop}[Gas expansion of Boltzmann weights]
For every finite $\Lambda\subset\lat$,
\begin{equation}\label{eq:gasexp}
\eee^{-\beta H_\Lambda}\;=\;\unit +\sum_{n=1}^{\infty} \frac{1}{n!} \sum_{\substack{(\B_1,\ldots,\B_n)\\ \B_i \in\pro^\Phi_\Lambda}}\prod_{i=1}^n\xi_{\B_i}\,\prod_{1\le i<j\le n} \one{\B_i\sim \B_j}\;, 
\end{equation}
with 
\begin{equation}\label{eq:gasexp.1}
\xi(\B)\;=\; \sum_{\B'\subset\B} (-1)^{\card{\B'\setminus \B}} \,\eee^{-\beta H_{\B'}}\;.
\end{equation}
\end{prop}
This expansion is the genesis of the different gas expansions used in the paper.  It involves  operator-valued fugacities $\xi$ defined by Hamiltonians that are only partially turned-on.  Notice that
\begin{equation}
\supp\bigl(\xi(\B)\bigr) \;=\; \bigcup\bigl\{X\in \B\bigr\}\;=\; \underline\B\;.
\end{equation}

Similar expansions, albeit subject to summability conditions, hold for the Hamiltonians $\overline H_\Lambda$, $H_\Lambda^\Theta$ and $H_\Lambda^{\rm per}$ [see \eqref{eq:rr.qq.1.4.0}, \eqref{eq:rr.qq.1.4.1}--\eqref{eq:rr.qq.1.4.2} and \eqref{eq:rr.qq.1.4}].  We shall not study optimal conditions for such gas expansions to be well defined, focusing rather in the (more restrictive) conditions for the convergence of the associated cluster expansions.

\subsection{{\bf Gas expansions for partition functions and expectations}}
 
The gas expansion of the partition function with free boundary conditions \eqref{eq:rr.qq.3} is obtained applying $\tr_\Lambda$ throughout \eqref{eq:gasexp}--\eqref{eq:gasexp.1} and making use of the factorization property (T3) of the trace.  In consistency with our previous notation we denote, for a finte family of bonds $\B$,
\begin{equation}
Z_{\B}\;=\; \tr_\Lambda\bigl(\eee^{-\beta H_\B}\bigr) \quad,\quad Z_\Lambda\equiv Z_{\B_\Lambda}\;.
\end{equation}

\begin{cor}
For every finite $\Lambda\subset\lat$,
\begin{equation}\label{eq:gasexp}
Z_\Lambda\;=\;1+\frac{1}{n!}\sum_{n=1}^{\infty} \sum_{\substack{(\B_1,\ldots,\B_n)\\ \B_i \in\pro_\Lambda}}\prod_{i=1}^n\rho_{\B_i}\,\prod_{1\le i<j\le n} \one{\B_i\sim \B_j}\;, 
\end{equation}
with
\begin{equation}\label{eq:gasexp.1}
\rho(\B)\;=\; \sum_{\B'\subset\B} (-1)^{\card{\B'\setminus \B}} \,Z_{\B'}\;=\; \tr_{\underline\B} \bigl(\xi(\B)\bigr)\;.
\end{equation}
\end{cor}
Similar expansions, subjected to the convergence conditions mentioned above, hold in the presence of boundary conditions, for the partition functions $Z_\Lambda(\Theta)$ and $ Z_\Lambda^{\beta,\mathrm per}$.  The respective fugacities ---denoted $\rho^\Theta$ and $\rho^{\rm per}$--- are also of the form \eqref{eq:gasexp.1} but using the partition functions $Z_\B(\Theta)$ and $Z^{\beta,{\mathrm per}}_\B$ for, respectively, $\B\in \B_{\partial\Lambda}$ and $\B\in \B^{\mathrm per}_{\partial_{\underline a}}$.  The respective polymers are denoted $\pro_{\partial\Lambda}$ and $\pro^{\mathrm per}_{\partial_{\underline a}}$.
\medskip

Gas expansions are also useful to describe the numerators of the expectations $\pi_\Lambda(A)$ defining finite-volume Gibbs states [\eqref{eq:rr.qq.2}--\eqref{eq:rr.qq.5}].  The corresponding expression is obtained as follows.  First, we notice that, as the sum in the right-hand side of \eqref{eq:gasexp} involves different families $\B_i$, we can replace 
$1/n! \sum_{(\B1,\ldots,\B_n)}$ by $\sum_{\{\B1,\ldots,\B_n\}}$. Second, we exploit the commutativity of the operator-valued fugacities $\xi(\B_i)$.  Third, we classify the families $\B_i$ according to whether they intersect $X_0:=\supp A$ or not.   In this way we obtain, from \eqref{eq:gasexp}, 
\begin{eqnarray}\label{eq:gasexp}
A\,\eee^{-\beta H_\Lambda}
 &=& A+\sum_{n=1}^{\infty} \sum_{\substack{\{\B_1,\ldots,\B_n\}\\ \B_i \in\pro_\Lambda}}A\,\prod_{i=1}^n\xi_{\B_i}\,\prod_{1\le i<j\le n} \one{\B_i\sim \B_j}\nonumber\\
&=& A\biggl\{1+\sum_{k\ge 1} \sum_{\substack{\{\B_1,\ldots,\B_k\}\\ \B_i \in\pro_\Lambda\\ \B_i\not\sim X_0}} \prod_{i=1}^k\xi_{\B_i}\,\prod_{1\le i<j\le k} \one{\B_i\sim \B_j} 
\nonumber\\
&& \qquad \times\ 
\biggl[1+\sum_{\ell\ge 1}\sum_{\substack{\{\B_{k+1},\ldots,\B_{k+\ell}\}\\ \B_i \in\pro_\Lambda\\ \B_i\sim X_0}} \prod_{u=k+1}^{k+\ell} \xi_{\B_j}\,\prod_{\substack{1\le u \le \ell\\ 1\le i \le n} }\one{\B_{k+u}\sim \B_i}\biggr]\biggr\}
\end{eqnarray}
The sum in the second line amounts to a sum over all families $\B\not\sim X_0$ of a single factor $\xi(\B)$ which is uniquely determined by its maximally connected decompositions $\{\B_1,\ldots,\B_k\}$.  The sum of the last line is the gas decomposition of the Boltzmann factor for the bonds compatible with $X_0$ and with the families of bonds in the previous factor.  The final expression, and its traced-out version, is summarized in the following.
 
\begin{prop}\label{theo:rr.tan.1.1}
Let $A\in\alg_{X_0}$ and $\Lambda\sset\lat$ with $X_0\subset\Lambda$.  Then, 
\begin{equation}\label{eq:gar.1}
A\,\eee^{-\beta H_\Lambda} \;=\; \sum_{\substack{\B\subset\B_\Lambda\\ \B\cup X_0\,{\rm connected}\\ \B\not\sim X_0}} A\,\xi(\B)\, \eee^{-\beta H_{\Lambda\setminus (X_0\cup\underline\B)}}
\end{equation}
Hence, taking the trace to both sides,
\begin{equation}\label{eq:quan8.ave}
\pi_\Lambda(A)\;=\; \sum_{\substack{\B\subset\B_\Lambda\\ \B\cup X_0\,{\rm connected}\\ \B\not\sim X_0}} K_\Lambda(A,\B) \,\frac{Z_{\Lambda\setminus (X_0\cup\underline\B)}}{Z_\Lambda}
\end{equation}
with
\begin{equation}\label{eq:quan8.ave.20}
 K_\Lambda(A,\B)  \;:=\; \sum_{\widetilde\B\subset\B_\Lambda}
(-1)^{\card{\B\setminus\widetilde\B}} \,
\tr_\Lambda \Big( A\,\eee^{-\beta H_{\widetilde\B}}\Bigr)\;.
\end{equation}
The case $\B=\emptyset$ is included in the sum \eqref{eq:gar.1} and, for the sake of compactness, we use the natural conventions $\xi(\emptyset)=\unit$ and $Z_\emptyset=1$.
\end{prop}
%

Modulo summability conditions, expressions analogous to \eqref{eq:gar.1} and \eqref{eq:quan8.ave} hold for the Hamiltonians $H_\Lambda^\Theta$ and $H_\Lambda^{\rm per}$ and the associated expectations $\pi_\Lambda(A \mid \Theta)$ and
$\pi^{\beta\,{\mathrm per}}_\Lambda(A) $.  These expressions involve fugacities and partition functions with the corresponding superscripts $\Theta$ and ``per".

\subsection{{\bf General cluster expansion}}

In general terms, cluster expansions are the formal power series obtained by expanding the logarithm of \eqref{eq:gengas} in powers of the fugacities $E(B_i)$.  
Such multivariate formal power series are also known as \empbf{polymer expansions}~\cite{GK71}. The expression of its terms was already determined in the original work by Mayer (see, for example, \cite[Proposition V.7.2]{Simon93} or \cite[Section 5.3]{FV17}).  It is  summarized in the next definition and theorem.  

\begin{defin} Consider a set $\mathcal S$ endowed with a hard-core compatibility relation ``$\sim$".  
\begin{itemize}
\item[(i)] The \empbf{incompatibility graph} of a sequence $(B_1,\ldots, B_n)\subset \mathcal S$ 
is the graph $\mathbb{G}(B_1,\ldots,B_n)$ vertex set $\{1,\ldots,n\}$ and edge set 
\[\left\{\{i,j\}:B_i\nsim B_j,0\le i<j\le n\right\}.\]
\item[(ii)] The sequence $(B_,,\ldots, B_n)$ is a \empbf{cluster} if it is connected or, equivalently, if $\mathbb{G}(B_1,\ldots,B_n)$ is a connected graph.
\end{itemize}
\end{defin}

\begin{defin} The \empbf{cluster expansion} corresponding to the gas expansion \eqref{eq:gengas} is the formal power series in the fugacities $E(B)$ obtained by taking the logarithm ---in the sense of composition of formal power series--- of the power series \eqref{eq:gengas}.
\end{defin}

\begin{thm}\label{theo:cl1} The cluster expansion for the general gas expansion \eqref{eq:gengas} is
\begin{equation}\label{eq:polmect10.-1}
\log F(\mathcal S)\;=\;\sum_{n=1}^{\infty}\frac{1}{n!}\sum_{\substack{(B_1,\ldots,B_n)\\ B_i\subset \mathcal S\\ B_i \;{\rm connected}}}\omega_n^T(B_1,\ldots,B_n)\,E(B_1)\cdots E(B_{n})
\end{equation}
with
\begin{equation}\label{eq:polmect12}
\omega_n^T(B_1,\ldots,B_n)\;=\;\left\{\begin{array}{cl}
1 & \text{if }\, n=1\\ [6pt]
\sum\limits_{\substack{G\subset\mathbb{G}(B_1,\ldots,B_n)\\G\, \mathrm{conn. spann.}}} (-1)^{|E(G)|} & \text{if }\,n>1 \text{ and }\mathbb{G}(B_1,\ldots,B_n)\text{ connected} \\ [6pt]
0 & \text{if } \mathbb{G}(B_1,\ldots,B_n)\, \text{ not connected}
\end{array}\right..
\end{equation}
where $G$ ranges over all  connected spanning subgraphs of $\mathbb{G}(B_1,\ldots,B_n)$. 
\end{thm}
Two comments are in order. 
\begin{rem}
The fugacities $E(B)$ in \eqref{eq:polmect10.-1} need not be commuting objects, in which case the order of the product is given by the order in the sequence $(B_1,\ldots, B_n)$.  
\end{rem}
\begin{rem}
The set $\mathcal S$ in \eqref{eq:polmect10.-1} need not be finite but, even if it is, the sum on the right-hand side involves infinitely many terms.  This is because, as the $B_i$ are not mutually compatible, there can be an arbitrary number of repetitions of each set.
\end{rem}

\subsection{{\bf Cluster expansions for free energies}}
The basic application of cluster expansions is as developments for the free energy.

\begin{thm}\label{thm:polmect1.1} In the sense of formal power series, the logarithm of the expansion \eqref{eq:gasexp} takes the form
\begin{equation}\label{eq:polmect10.1}
\log Z_\Lambda\;=\;\sum_{n=1}^{\infty}\frac{1}{n!}\sum_{n=1}^{\infty}\sum_{\substack{(\B_1,\ldots,\B_n)\\ \B_i \in\pro_\Lambda}}\omega_n^T(\B_1,\ldots,\B_n)\,\rho_{\B_1}\ldots \rho_{\B_{n}}
\end{equation}
with $\omega_n^T(\B_1,\ldots,\B_n)$ defined in \eqref{eq:polmect12}.
\end{thm}

Analogous expressions hold for $\log Z^\#_\Lambda$ with fugacities $\rho^\#_\B$ and polymers chosen from families $\pro^\#_\Lambda$ of bonds in $\B_{\partial\Lambda}$ and $\B^{\mathrm per}_{\partial_{\underline a}}$, respectively for $\#=\Theta, {\rm per}$.

To see more clearly the handling of cluster expansions, it is convenient to adopt a more compact notation.  Let $\mathbb C_{\B}$ denote the set of clusters formed out of bonds in $\B$:
\begin{equation}
\mathbb C_{\B}\;=\; \bigl\{(\B_1,\ldots,\B_n) : n\in \mathbb N_{>0}, \B_i\subset \B, \B_i \mbox{ connected}, \mathbb{G}(\B_1,\ldots,\B_n)\text{ connected} \bigr\}\;;
\end{equation}
then \eqref{eq:polmect10.1} can be written as 
\begin{equation}
\log Z_\Lambda\;=\;\sum_{\mathcal C\in \mathbb C_{\B_\Lambda}} \Omega^T(\mathcal C)
\end{equation}
with
\begin{equation}
\Omega^T(\B_1,\ldots,\B_n)\;=\; \frac{1}{n!} \,\omega_n^T(\B_1,\ldots,\B_n) \,\rho_{\B_1}\ldots \rho_{\B_{n}}\;.
\end{equation}
Let us emphasize that clusters are \emph{ordered} sequences of polymers.  With this compact notation Theorem \ref{thm:polmect1.1} generalizes to arbitrary boundary conditions as follows.
\begin{thm}\label{thm:polmect1.1.1} In the sense of formal power series, 
\begin{equation}\label{eq:polmect10.1.1.A}
\log Z^\#_\Lambda\;=\; \sum_{\mathcal C\in \mathbb C_{\B^\#_\Lambda}} \Omega^{T,\#}(\mathcal C)
\end{equation}
with
\begin{equation}
\Omega^{T,\#} (\B_1,\ldots,\B_n)\;=\; \frac{1}{n!} \,\omega_n^T(\B_1,\ldots,\B_n) \,\rho^\#_{\B_1}\ldots \rho^\#_{\B_{n}}
\end{equation}
and $\B^\#_\Lambda = \B_\Lambda\,,\B_{\partial\Lambda}\,,
\B^{\mathrm per}_{\partial_{\underline a}}$
for  $\#=\emptyset,\Theta, {\rm per}$.

\end{thm}
\medskip

\section{Results II: Convergence criteria}\label{ssec:rr.tan.10}

\subsection{{\bf General criterion}}

The convergence of the series \eqref{eq:polmect10.-1}--\eqref{eq:polmect12} has been addressed through a number of approaches that includes the use of (i) tree-graph inequalities \cite{Bry84}, (ii) inductive proofs \cite{KP86,Dob96,Mir00,PU09} and (iii) tree-graph identities \cite{PF07,PY17,FN19,fia20}.  Approach (ii) has been shown to be basically equivalent to an improved analysis of KS equations \cite{BFP10}.  Approach (iii) is a refinement of approach (ii) and it is the one leading to the strongest results.  We adopt the latter here.  We shall use boldface to denote functions on polymers, for instance $\bbd\rho=\{\rho_Y:Y\in\pro\}$

Convergence is studied at the level of the \emph{pinned cluster expansion}
\begin{equation}\label{eq:polmect10.1.1}
\Sigma_{B_0}(\bbd{E})\;=\;\sum_{n=1}^{\infty}\frac{1}{n!}\sum_{\substack{(B_1,\ldots,B_n)\\ B_i\in\pro}}\omega_n^T(B_0,B_1,\ldots,B_n)\,E(B_1)\cdots E(B_{n})
\end{equation}
which is controlled by the series obtained by termwise bounding by absolute values:
\begin{equation}\label{eq:polmect10.1.2.bis}
\card{\Sigma}_{B_0}(\bbd{\lambda})\;=\; \sum_{n=1}^{\infty}\frac{1}{n!}\sum_{\substack{(B_1,\ldots,B_n)\\ B_i\in\pro}}\card{\omega_n^T(B_0,B_1,\ldots,B_n)}\,\lambda_{\B_1}\cdots \lambda_{\B_{n}}
\end{equation}
with $\lambda_{\B_i}\in\mathbb{R}_+$.  As this last series has positive terms, its convergence amounts to boundedness of the partial sums; a fact relatively simple to determine without recourse to Banach-space fixed points or other mathematically sophisticated techniques.   The convergence of \eqref{eq:polmect10.1.2.bis} implies the absolute convergence of \eqref{eq:polmect10.1.bis} uniformly in the polydisc 
\begin{equation}
\mathcal{D}(\bbd\lambda)\;=\; \bigl\{\card{\rho_B}\le \lambda_B:B\in\pro\bigr\}\;.  
\end{equation}

Notice that for the free-energy expansion \eqref{eq:polmect10.1}, the pinned expansion \eqref{eq:polmect10.1.1} becomes
\begin{equation}\label{eq:polmect10.1.bis}
\Sigma_{\B_0}(\bbd{\rho})\;=\;\sum_{n=1}^{\infty}\frac{1}{n!}\sum_{n=1}^{\infty}\sum_{\substack{(\B_1,\ldots,\B_n)\\ \B_i \in\pro}}\omega_n^T(\B_0,\B_1,\ldots,\B_n)\,\rho_{\B_1}\ldots \rho_{\B_{n}}\;.
\end{equation}
Free energies are related to the finite-region pinned expansions $\Sigma^\Lambda_{\B_0}(\bbd{\rho})$ which are defined by restricting sums to polymers $B_i\in\pro_\Lambda$ and, more generally, introducing boundary conditions:
\begin{equation}\label{eq:polmect10.1.bis.bc}
\Sigma^{\Lambda,\#}_{\B_0}(\bbd{\rho})\;=\;\sum_{n=1}^{\infty}\frac{1}{n!}\sum_{n=1}^{\infty}\sum_{\substack{(\B_1,\ldots,\B_n)\\ \B_i \in\pro^\#_\Lambda}}\omega_n^T(\B_0,\B_1,\ldots,\B_n)\,\rho^\#_{\B_1}\ldots \rho^\#_{\B_{n}}\;.
\end{equation}
with $\pro^\#_\Lambda$ formed by families of bonds respectively in
and $\B^\#_\Lambda = \B_\Lambda\,,\B_{\partial\Lambda}\,,
\B^{\mathrm per}_{\partial_{\underline a}}$
for  $\#=\emptyset,\Theta, {\rm per}$.

In the sense of formal power series, these pinned expansions are derivatives of the free energy:
\begin{equation}\label{eq:rr.bul.1}
\frac{\partial}{\partial \rho_{\B_0}}\log Z^\#_\Lambda(\bbd{\rho})\;=\; \Sigma^{\Lambda\#}_{\B_0}(\bbd{\rho})
\end{equation}
for $\B_0\subset\B_\Lambda$.
This makes $\bbd\Sigma^\Lambda$ the right objects for telescopic arguments in which polymers are incorporated one at a time.  This is made explicit by the relation 
\begin{equation}\label{eq:rr.bul.1.1}
\frac{\partial}{\partial \rho_{\B_0}}\log Z^\#_\Lambda\;=\; \frac{1}{ Z^\#_\Lambda}\,\frac{\partial  Z^\#_\Lambda}{\partial \rho_{\B_0}}\;=\; 
\frac{ Z^\#_{\B_\Lambda\setminus \B_0}}{ Z^\#_\Lambda}
\end{equation}
(notice that $Z^\#_\Lambda$ is a linear function of each $\rho_{\B}$).  These identities become functional identities upon convergence.  

The following theorem states the strongest convergence criterion available for the pinned expansion \eqref{eq:polmect10.1.1}.

\begin{thm}\label{theo:rr.tan.06}
Consider the function $\boldsymbol{\varphi}:[0,+\infty)^{\mathcal{P}} \longrightarrow [0,+\infty]^{\mathcal{P}}$ defined by
\begin{equation}\label{eq:quan26}
\varphi_{B_0}(\boldsymbol{\mu})\;=\;
1+\sum_{n\ge1}\frac{1}{n!}\sum_{\substack{(B_1,\ldots,B_n)\\ B_i\in\pro}}\mu_{B_1}\cdots \mu_{B_n}
\prod_{i=1}^n\one{B_0\nsim B_i}\prod_{1\le k< \ell\le n}\one{B_k\sim B_\ell}\;.
\end{equation}
If $\boldsymbol{\lambda}\in[0,+\infty)^{\mathcal{P}}$ satisfies 
\begin{equation}\label{eq:quan23.1}
\lambda_{B}\;\le\; \frac{\mu_{B}}{\varphi_{B}(\boldsymbol{\mu})}
\end{equation}
for each $B\in\mathcal{P}$, for some $\boldsymbol{\mu}\in[0,+\infty)^{\mathcal{P}}$, then 
the following holds uniformly in the polydisk $\mathcal{D}(\bbd\lambda):=\bigl\{\bbd\rho: \card{\rho_B}\le \lambda_B , B\in \pro\bigr\}$:
\begin{itemize}
\item[(a)] The pinned expansion \eqref{eq:polmect10.1.1} converge absolutely.

\item[(b)] For each $B\in\mathcal{P}$,
\begin{equation}\label{eq:mmm.1}
\card{\rho(B)}\,\card{\Sigma_B(\bbd{\rho})}\;\le\; \mu(B)\;.
\end{equation}
\end{itemize}

\end{thm}

The application of this theorem involves, of course, the determination of functions $\bbd\mu$ maximizing the right-hand side of \eqref{eq:quan23.1}.
 There exist three different proofs of part (a) of this theorem~\cite{PF07,FN19,fia20}.  Part (b) is proven in \cite{PF07,FN19} where, in fact, a whole sequence of improvements of this bound is presented. These bounds, and one of the key steps in the proof of the criterion \eqref{eq:quan23.1} are explained in Appendix \ref{app.te}.

For ratios of partition functions, the previous theorem yields the bound

\begin{cor}\label{cor:bio.1}
The convergence criterion \eqref{eq:quan23.1} implies the absolute convergence of the free-energy expansion \eqref{eq:polmect10.1}.  Furthermore, it yields the bound
\begin{equation}\label{eq:rr.bul.3}
\card{\frac{ Z_{\B_\Lambda\setminus \B}}{ Z_\Lambda}}\;\le\;\frac{\mu_\B}{\lambda_\B}
\end{equation}
for each $\B\in\pro_\Lambda$. The bond can be further improved by the following sequence of inequalities.  For $\B\in\pro_\Lambda$
\begin{equation}\label{eq:rr.bul.2}
\card{\log \frac{ Z_\Lambda}{ Z_{\B_\Lambda\setminus \B}}} \;\le\;
-\varphi_\B(\bbd\mu)\,\log(1-\lambda_\B)\;\le \log\bigl(1+\lambda_\B\,\varphi_\B(\bbd\mu)\bigr)\;\le\; \lambda_\B\,\varphi_\B(\bbd\mu)\;.
\end{equation}
\end{cor}
It is not immediately clear whether the analogous of this last corollary holds for systems with non-empty boundary conditions.  The reason is that the absolute convergence of the pinned expansion \eqref{eq:polmect10.1.bis} does not immediately imply convergence of the (free-energy) cluster expansions in the presence of boundary conditions, because there is no clear relation between the norms of the fugacities $\rho^\Theta_\B$ and that of the free fugacities $\rho_\B$.  We will return to this issue further down, when presenting convergence conditions that are insensitive to boundary conditions.

\subsection{{\bf Criterion for polymers with geometric support}}
Criterion \eqref{eq:quan23.1} is very performant if each polymer has only a few incompatible ones around it.  For cases with large, potentially infinite incompatibility neighborhoods ---as in our case--- there are more efficient weaker sufficient conditions.  Let us focus in the case ---of interest here--- in which each polymer $B$ has an associated geometric support $\underline B$ and compatibility corresponds to disjointness of supports.  In this situation, a convenient sufficient convergence criterion is obtained by majorizing $\bbd\varphi$ iin \eqref{eq:quan23.1} by the function 

\begin{equation}
\widetilde \varphi_{B_0}\;=\; \prod_{x\in \underline B_0}\Bigl[1+\sum_{\underline B\ni x}\mu_{B}\Bigr]\;,
\end{equation}
which is obtained by bounding
\begin{equation}
\one{B_0\nsim B_k}\,\one{B_0\nsim B_\ell}\,\one{B_k\sim B_\ell}\; \le\;
\one{B_0\nsim B_k}\,\one{B_0\nsim B_\ell}\,\one{\underline B_k\cup \underline B_0\neq \underline B_\ell\cup \underline B_0}\;.
\end{equation}
The resulting criterion is a slight improvement of a criterion due to Gruber and Kunz~\cite{GK71}, originally proven in~\cite{PF07} and, with an alternative approach in~\cite{BFP10}.

\begin{cor}\label{corollary1}
A sufficient criterion for \eqref{eq:quan23.1} is the existence of $\boldsymbol{\lambda}\in[0,+\infty)^{\mathcal{P}}$ such that
\begin{equation}\label{eq:quan23.r1}
\lambda_{B_0}\;\le\; \frac{\mu_{B_0}}{\prod_{x\in \underline B_0}\Bigl[1+\sum_{\underline B\ni x}\mu_{B}\Bigr]}\;.
\end{equation}
for each $B_0\in\mathcal{P}$, for some $\boldsymbol{\mu}\in[0,+\infty)^{\mathcal{P}}$,
\end{cor}
It is clear that the bounds \eqref{eq:rr.bul.2} hold also with $\bbd{\widetilde\varphi}$ in the place of $\bbd\varphi$.

Condition \eqref{eq:quan23.1} shows that the ratios $\mu_B/\lambda_B$ cannot be smaller than one.  This justifies changing from $\bbd\mu$ to exponential weights  $a(B)\ge 0$ through the reparametrization
\begin{equation}
\mu_B=\lambda_B\,\eee^{a(B)}
\end{equation}
which is, in fact, the one used in early convergence criteria~\cite{Bry84,KP86,Dob96,Mir00}.  With this, the previous corollary yields the following slightly weaker criterion.

\begin{cor}[Improved GK-criterion]\label{corollary1.1}
A sufficient criterion for \eqref{eq:quan23.r1}, and hence for \eqref{eq:quan23.1}, is the existence of $\boldsymbol{\lambda}\in[0,+\infty)^{\mathcal{P}}$ such that 
\begin{equation}\label{eq:rr.convergence.1}
\sup_{x\in \underline B_0}\sum_{\substack{B\in\pro\\x\in \underline B}}\lambda_{B}\eee^{a(B)}\le \eee^{a(B_0)/|B_0|}-1
\end{equation}
for each $B_0\in\pro$.
In this case, 
\begin{equation}\label{eq:rr.convergence.1.bis}
\card{\Sigma}_B\;\le\; \eee^{a(B)}
\end{equation}
for each $B\in\pro$.
\end{cor}
In terms of this criterion, the analogous of Corollary \ref{cor:bio.1} reads as follows.
\begin{cor}\label{cor:bio.2}
The convergence criterion \eqref{eq:rr.convergence.1} applied to the free-energy expansion \eqref{eq:polmect10.1} yields the bounds
\begin{equation}
\label{eq:rr.bul.3.bis}
\card{\frac{ Z_{\B_\Lambda\setminus \B}}{ Z_\Lambda}}\;\le\; \eee^{a(\B)}
\end{equation}
and
\begin{equation}\label{eq:rr.bul.2.bis}
\card{\log \frac{ Z_\Lambda}{ Z_{\B_\Lambda\setminus \B}}} \;\le\; \log\bigl(1+\lambda_\B\,\eee^{a(\B)}\bigr)\;\le\; \lambda_\B\,\eee^{a(\B)}\;.
\end{equation}
\end{cor}


\section{Results III: Convergence criteria in terms of interactions}\label{ssec:rr.tan.10.int}

\subsection{{\bf Improved GK criterion in terms of interactions}}

To exploit the preceding criteria in concrete applications it is necessary to establish some characterization of $\mathcal{D}(\bbd\lambda)$, or some non-empty region therein, in terms of $\beta$ and the interaction terms $\Phi(X)$.  The link is provided by the following bound, basically shown by Park \cite{Park82}
\begin{lem}\label{Lemma2}
\begin{equation}\label{eq:rr.oeg.trad}
\left\|\xi_\B\right\|\;\le\; \prod_{X\in\B}\Bigl[ \eee^{\beta \|\Phi(X)\|} -1\Bigr]\;.
\end{equation}
As a consequence, 
\begin{equation}\label{eq:rr.bound}
|\rho^\#_\B|\;\le\; \prod_{X\in\B}\Bigl[ \eee^{\beta \|\Phi(X)\|} -1\Bigr]
\end{equation}
for $\#=\theta, {\rm per}, \emptyset$.
\end{lem}
The proof is presented below for completeness.     
The bound \eqref{eq:rr.bound} is optimal for classical interactions because, if the operators $\Phi(X)$ commute,
\begin{equation}\label{eq:conf.ibc}
\rho_\B \;=\; \sum_{\B'\subset\B} (-1)^{\card{\B'\setminus\B}}\, \prod_{X\in\B'} \eee^{-\beta\Phi(X)} \;=\; \prod_{X\in\B} \Bigl(\eee^{-\beta\Phi(X)} -1 \Bigr)\;.
\end{equation}
Furthermore, for the classical case, if the interaction is non-negative, the bound \eqref{eq:conf.ibc} can be improved to $\prod_{X\in\B}\bigl(1-\eee^{-\beta \,\Phi(X)} \bigr)\;\le\; \lambda_{\B}$.

Denoting 
\begin{equation}\label{eq:rr.bound.v}
V_\B\;=\; \prod_{X\in\B}\Bigl[ \eee^{\beta \|\Phi(X)\|} -1\Bigr]\;,
\end{equation}
we replace \eqref{eq:polmect10.1.2.bis} by its majoration
\begin{equation}\label{eq:polmect10.1.2.bissquare.v}
\bigl|\widehat\Sigma\bigr|_{B_0}(\bbd{V})\;=\; \sum_{n=1}^{\infty}\frac{1}{n!}\sum_{\substack{(B_1,\ldots,B_n)\\ B_i\in\pro}}\card{\omega_n^T(\B_0,\B_1,\ldots,\B_n)}\,V_{\B_1}\cdots V_{\B_{n}}\;.
\end{equation}
The convergence of this expansion implies convergence of pinned and cluster expansions for \emph{all} boundary conditions.  


At this point, we adopt the convergence condition \eqref{eq:rr.convergence.1} and, as customary in standard cluster-expansion treatments, we choose 
\begin{equation}\label{eq:choice}
a(\B)\;=\;a\card{\underline\B}\quad,\quad a>0\;.
\end{equation}
The immediate advantage of this choice is that the optimization inherent to condition \eqref{eq:rr.convergence.1} involves the single parameter $a$.  This makes numerical determination much simpler and often simple calculus consideration suffice.  
With this choice, the convergence criterion takes the following form. 

Applying the condition \eqref{eq:rr.convergence.1} with the choice \eqref{eq:choice} we obtain the following convergence criterion. Let
\begin{equation}\label{eq:quan34.bis}
T_a(\beta)\,:= \, \sup_x\sum_{n\ge 1}  \sum_{\substack{\{X_1,\ldots, X_n\}\\X_i\in\bonds ,x\in X_1\\\mathbb G(X_1,\ldots, X_n) \,\mathrm{connected}}} 
\prod_{i=1}^n \left(\eee^{\beta\|\Phi(X_i)\|}-1\right)\eee^{a|X_i|}\;.
\end{equation}

\begin{cor}[GK criterion in terms of interactions I]   \label{eq:cor1}
If there exist $a>0$ such that 
\begin{equation}\label{eq:rr.1}
T_a(\beta) \;\le\; \eee^a-1
\end{equation}
then the following holds.
\begin{itemize}
\item[(a)] The expansion \eqref{eq:polmect10.1.2.bissquare.v} converges and, hence,
the pinned cluster expansions \eqref{eq:polmect10.1.bis.bc} and the free-energy cluster expansions 
\eqref{eq:polmect10.1.1.A}
converges absolutely and uniformly in $\Lambda$ for $\#=\theta, {\rm per}, \emptyset$.  Furthermore,
\begin{equation}\label{eq:polmect10.1.2.bissquare.v.v}
\bigl|\widehat\Sigma\bigr|_{B_0}(\bbd{V})\;\le\; \eee^{a \card{\underline\B_0}};.
\end{equation}
\item[(b)]  The following bounds hold
\begin{equation}\label{eq:rr.2}
 \left|\frac{Z^\#_{\B_\Lambda\setminus\B_0}}{Z^\#_{\Lambda}}\right| 
  \;\le\; \eee^{a \card{\underline\B_0}}
\end{equation}
and
\begin{equation}\label{eq:rr.2.bis}
 \left|\log \frac{Z^\#_{\Lambda}}{Z^\#_{\B_\Lambda\setminus \B_0}}\right| \;\le\; \prod_{X\in\B_0} \left(\eee^{\beta\|\Phi(X)\|}-1\right)\eee^{a|X|}
\end{equation}
for every finite $\Lambda$, for every $\B_0\subset\B_\Lambda$, for $\#=\theta, {\rm per}, \emptyset$.
\item[(c)]  The site-pinned majorizers
\begin{equation}\label{eq:site-pinned}
\bigl|\widehat\Sigma\bigr|_{x}(\bbd{V})\;=\; \sum_{n=1}^{\infty}\frac{1}{n!}\sum_{\substack{(\B_1,\ldots,\B_n)\\ \B_i\in\pro\\x\in\cup_i\underline{\B}_i}}\card{\omega_n^T(\B_1,\ldots,\B_n)}\,V_{\B_1}\cdots V_{\B_{n}}\;.
\end{equation}
satisfies
\begin{equation}\label{eq:site-pinned.b}
\bigl|\widehat\Sigma\bigr|_{x}(\bbd{V})\;\le\; \eee^a-1\;.
\end{equation}
\end{itemize}
\end{cor}
Notice that, in obtaining \eqref{eq:quan34.bis}--\eqref{eq:rr.1} we bounded
$\card{\cup_i X_i}\le \sum_i \card{X_i}$.  In principle, the parameter $a$ can depend on $\beta$.  In practice, however, it is often determined without regard to $\beta$ and the latter is subsequently adjusted.

\subsection{{\bf A simpler restatement of GK criterion}}

The stumbling block in the application of the criterion of Corollary \ref{eq:cor1} is the evaluation of the infinite sum $T_a(\beta)$ in \eqref{eq:quan34.bis}.  For general weights $W(X)$ this sum is of the form
\begin{equation}\label{eq:min.1.1}
T_a(\beta) \;=\; \sup_x\sum_{X_0\in\bonds,X_0\ni x} W(X_0)\, T(X_0)
\end{equation}
with
\begin{equation}\label{eq:every.1}
T(X_0)\;=\; 1+ \sum_{n\ge 1}  \sum_{\substack{\{X_1,\ldots, X_n\}\\X_i\in\bonds , X_i\neq X_0}} 
\one{\mathbb G(X_0,X_1,\ldots, X_n) \,\mathrm{connected}}
\prod_{i=1}^n W(X_i)\;.
\end{equation}
The best way to perform this last sum is by reordering it in terms of trees.  Indeed, the well known classification of graphs in partition schemes (see Appendix \ref{app.te}) implies that we can define (in fact, in many different ways) a map that to each connected graph $\mathbb G$ with vertices $(X_0,X_1,\ldots, X_n)$ associates a unique planar spanning tree $\mathcal T(\mathbb G)\subset \mathbb G$ with root at $X_0$.  
\begin{equation}
\one{\mathbb G(X_0,X_1,\ldots, X_n) \,\mathrm{connected}}\;=\; \one{\mathcal T(G(X_0,X_1,\ldots, X_n)) \mbox{ spanning}}
\end{equation}
Replacing this in \eqref{eq:every.1}, we can then turn the tree condition into a sum over trees and a sum over labels leading to such trees: 
\begin{equation}\label{eq:every.2}
T(X_0)\;=\; 1+ \sum_{n\ge 1}  \sum_{\tau\in \mathbb T_n} \sum_{\substack{\{X_1,\ldots, X_n\}\\X_i\in\bonds , X_i\neq X_0}} 
\one{\mathcal T(\mathbb G(X_0,X_1,\ldots, X_n))=\tau}
\prod_{i=1}^n W(X_i)
\end{equation}
where $\mathbb T_n$ denots the set of planar rooted trees with $n$ non-root vertices.
As illustrated in Appendix \ref{app.te}, trees $\mathcal T(\mathbb G)$ obtained through partition schemes satisfy several order constraints on the vertex labels $X_i$.  These constraints can, perhaps, be exploited to obtain more refined estimations, but, for the sake of simplicity, here we retain the most basic one: the bonds labelling the vertices are all \emph{different}.  This leads to the following upper bound involving only a ``1st-generation" constraint.

\begin{prop}
Let us identify the vertices of a rooted  tree with an index $i=0,1,\ldots,n$, reserving $i=0$ for the root.  Each planar tree is determined by the number $s_i$ of children of each vertex $i$, identified by pair of indices $(i,1),\ldots,(i,s_i)$.  Then, 
\begin{equation}\label{eq:t01}
T(X_0)\;\le\; \widehat T(X_0)\;=:\; 1+ \sum_{n\ge 1} \sum_{\tau\in \mathbb T_n} \sum_{\substack{\{X_1,\ldots, X_n\}\\X_i\in\bonds , X_i\neq X_0}} 
\prod_{i=0}^n c_{s_i}\bigl(X_i,X_{(i,1)},\ldots, X_{(i,s_i)}\bigr)\prod_{j=1}^n W(X_j)
\end{equation}
with
\begin{equation}
c_{s_i}\bigl(X_i,X_{(i,1)},\ldots, X_{(i,s_i)}\;=\; \prod_{j=1}^n\one{X_{(i,j)}\cap X_i \neq\emptyset\,,\,X_{(i,j)}\neq X_i}
\prod_{1\le k<\ell\le n}\one{X_{(i,k)}\neq X_{(i,\ell)}}\;.
\end{equation}  
\end{prop}

At this point we can resort to the canonical way of summing tree expansions with positive weights (reviewed in Proposition \ref{prop:cast} in Appendix \ref{app.te}).  As trees with $n+1$ generations are inductively generated by placing $n$-generation trees on the leaves of first-generation trees, it is necessary and sufficient to bound the latter to generate an upper bound that guarantees convergence of the positive series.  This bound requires the existence of positive weights $\zeta$ such that for each $X$, the $X$-rooted first-generation trees corresponding to the right-hand side of \eqref{eq:t01}, 
\begin{equation}
\widehat T_1(X)\;=\; 1+ \sum_{s\ge 1}\sum_{\substack{\{X_1,\ldots, X_n\}\\X_i\in\bonds , X_i\neq X}} 
c_{s}\bigl(X,X_1,\ldots, X_s\bigr)\prod_{j=1}^n \zeta(X_j)
\end{equation}
satisfy
\begin{equation}\label{eq:opera.it}
W(X)\,\widehat T_1(X)\;\le\; \zeta(X)\;.
\end{equation}
Working iteratively in different generations, this bound extends then to the inequality $W(X)\,\widehat T_1(X)\le \zeta(X)$.
As,
\begin{equation}
\widehat T_1(X)\;=\;1 + \sum_{n\ge 1} \sum_{\substack{\{X_1,\ldots, X_n\}\\X_i\in\bonds , X_i\neq X\\ X_i\cap X\neq\emptyset} }
\prod_{j=1}^n \zeta(X_j)
\;=\; \prod_{\substack{Y\in\bonds ,\,Y\neq X\\Y\cap X\neq\emptyset}}\bigl[1+\zeta(Y)\bigr]\;,
\end{equation}
Proposition \ref{prop:cast}  yields the following bound.

\begin{prop}\label{prop.xuan.1} If the weights $W$ satisfy 
\begin{equation}\label{eq:min.2}
W(X)\;\le\; \frac{\zeta(X)}{ \displaystyle\prod_{\substack{Y\in\bonds ,\,Y\neq X\\Y\cap X\neq\emptyset}}\bigl[1+\zeta(Y)\bigr]}
\end{equation}
for some $\zeta:\bonds \to (0,\infty)$, then
\begin{equation}\label{eq:min.3}
W(X)\,T(X)\;\le\;\zeta(X)
\end{equation}
for evey $X\in\bonds$.
 \end{prop}
 
To bound \eqref{eq:quan34.bis} we apply Proposition \ref{prop.xuan.1} for
 \begin{equation}\label{eq:min.1}
W(X)\;=\;  \left(\eee^{\beta\|\Phi(X)\|}-1\right)\eee^{a|X|}\;.
\end{equation}
From \eqref{eq:rr.1}, \eqref{eq:min.1.1} and \eqref{eq:min.3} we see that the function $\zeta$ should be related with the parameter $a$ by the identity
\begin{equation}\label{eq:min.5}
1+ \sup_x \sum_{X\in\bonds , X\ni x} \zeta(X)\;=\; \eee^a\;.
\end{equation}
We then obtain the following corollary of Corollary \ref{eq:cor1}.

\begin{cor}[GK criterion in terms of interactions II]   \label{eq:cor1.bis}
A sufficient condition for the absolute convergence of the pinned cluster expansion \eqref{eq:polmect10.1.bis} is the existence of a function $\zeta:\bonds \to (0,\infty)$ such that
\begin{equation}\label{eq:min.6}
\eee^{\beta\|\Phi(X)\|}-1\;\le\; \frac{\zeta(X)}{\displaystyle \Bigl[1+ \sup_y \sum_{\substack{Y\in\bonds\\ Y\ni y}} \zeta(Y)\Bigr]^{\card X}
\prod_{\substack{Y\in\bonds \\ Y\cap X\neq\emptyset\\Y\neq X}} \bigl[1+\zeta(Y)\bigr]}\;.
\end{equation}
Furthermore, the bounds \eqref{eq:rr.2} hold with the replacement \eqref{eq:min.5}.  Regarding \eqref{eq:rr.2.bis}, we have, from \eqref{eq:min.1} and \eqref{eq:min.2}, that
\begin{equation}\label{eq:rr.2.bis.bis}
 \left|\log \frac{Z^\#_{\Lambda}}{Z^\#_{\B_\Lambda\setminus \{X\}}}\right| \;\le\; 
 \frac{\zeta(X)}{ \prod_{\substack{Y\in\bonds \\Y\neq X}}\bigl[1+\zeta(Y)\bigr]}
\end{equation}
for each $X\in\bonds$, for $\#=\theta, {\rm per}, \emptyset$.  Bounds for $\left|\log \frac{Z^\#_{\Lambda}}{Z^\#_{\B_\Lambda\setminus \B_0}}\right|$ are telescopically given by products of the right-hand side for $X\in\B_0$.

\end{cor}

For the sake of computational simplicity, the right-hand side of \eqref{eq:min.6} can be successively bounded in the form
\begin{eqnarray}
\Bigl[1+ \sup_y \sum_{\substack{Y\in\bonds\\ Y\ni y}} \zeta(Y)\Bigr]^{\card X}
\prod_{\substack{Y\in\bonds \\ Y\cap X\neq\emptyset\\Y\neq X}} \bigl[1+\zeta(Y)\bigr] &\le&
\biggl[ \sup_y \prod_{\substack{Y\in\bonds \\ Y\ni y}} \bigl[1+\zeta(Y)\biggr]^{2\card X}\nonumber\\
&\le& \exp\biggl[2\card X \sup_y \sum_{\substack{Y\in\bonds\\ Y\ni y}} \zeta(Y)\biggr]\;.
\end{eqnarray}
These bound are not expected to be too pessimistic because usually the weights $\zeta(X)$ are rather small.  These bounds yield the following sufficient conditions for \eqref{eq:min.6}:
\begin{equation}\label{eq:min.6.a}
\eee^{\beta\|\Phi(X)\|}-1\;\le\; \frac{\zeta(X)}{\displaystyle \biggl[ \sup_y \prod_{\substack{Y\in\bonds \\ Y\ni y}} \bigl[1+\zeta(Y)\biggr]^{2\card X}}
\end{equation}
and
\begin{equation}\label{eq:min.6.b}
\eee^{\beta\|\Phi(X)\|}-1\;\le\; \zeta(X)\, \exp\biggl[-2\card X \sup_y \sum_{\substack{Y\in\bonds\\ Y\ni y}} \zeta(Y)\biggr]\;.
\end{equation}

\subsection{{\bf Applications}}

\subsubsection{{\bf Nearest-neighbor spin models}}

For simplicity, let us consider the translation- and rotation-invariant version.  Bonds are pairs of nearest-neighbor sites 
\begin{equation}
\bonds\;=\; \bigl\{ X=\{x,y\} : x,y \mbox{ nearest neighbors}\bigr\}
\end{equation}
and interactions are of the form
\begin{equation}
\Phi(\{x,y\})) \;=\;\Phi(\sigma_x,\sigma_y) 
\end{equation}
for some function $\Phi$ of the spin operators at nearest-neighbor sites.  In this situation, it is natural to choose constant $\zeta(X)=\zeta$.  Furthermore, if $\lat=\mathbb Z^d$, then for each $y\in\lat$ and each $X\in\bonds$,
\begin{eqnarray}
\card{\{Y\in \bonds: X\ni y\}} &=& 2d\\
\card{\{Y\in\bonds: Y\cap X\neq\emptyset\}} &=& 2(2d-1)\;.
\end{eqnarray}
Hence condition \eqref{eq:min.6} becomes
\begin{equation}\label{eq:min.7}
\eee^{\beta\|\Phi\|}-1\;\le\; \frac{\zeta}{\displaystyle \bigl[1+ 2d\,\zeta\bigr]^2
 \bigl[1+\zeta\bigr]^{4d-2}}\;.
\end{equation}
The optimal choice for $\xi$, namely the maximizer of the right-hand side, is
\begin{equation}
\widehat \zeta \;=\; \frac{3(1-2d)\pm\sqrt{68d^2-36d-7}}{8(d^2-1)}
\end{equation}

Table \ref{table:chi} summarizes numerical values obtained from \eqref{eq:min.7} with $\xi=\widehat\xi$  for different dimensions $d$. For comparison, the last column refers to the bound 
\begin{equation}
\beta\|\Phi\|\,\eee^{\beta\|\Phi\|}\;\le\;\frac{0.03}{d}\Bigl(1+ \frac{0.03}{d}\Bigr)
\end{equation}
derived from Park's work~\cite{Park82} (see Appendix \ref{sec:appendix}; note that $\eee^x-1\le x\,\eee^x$), itself a big improvement from the bound
\begin{equation}
\beta\,\eee^{\beta}\;\le\;\frac{1}{48\,d^2}
\end{equation}
obtained by Simon~\cite[Example 2, page 462]{Simon93} for the classical Ising model.  (Note that $\eee^x-1\le x\,\eee^x$.)

\begin{table}[h]
\begin{center}
\renewcommand{\arraystretch}{2}
\begin{tabular}{|c||c|c|c|}
\hline
$d$ & $\widehat\zeta$ & $\eee^{\beta\|\Phi\|}-1\le$ & Park \\
\hline\hline
2 & 0.087 & 0.029 & 0.015\\
\hline
3 & 0.054 & 0.018 &0.010\\
\hline
4 & 0.039 & 0.013 & 0.008\\
\hline
$\gg 1$ & $\frac{0.14}{d}\bigr[1+O(\frac1d)\bigr]$ & $\frac{0.032}{d} \bigr[1+O(\frac1d)\bigr]$ & $\frac{0.03}{d}\bigr[1+O(\frac1d)\bigr]$\\
\hline
\end{tabular}
\caption{$\beta$-radius of convergence of the cluster expansion for nearest-neighbor interactions}
\label{table:chi}
\end{center} 
\end{table}

\subsubsection{{\bf General spin models}}
To obtain an universal bound on $\beta$ we resort to the sufficient condition \eqref{eq:min.6.b} and the bounds
\begin{equation}
\eee^{\beta\|\Phi(X)\|}-1\;\le\; \beta\,\|\Phi(X)\|\, \eee^{\beta\|\Phi(X)\|} \;\le\; 
\beta\,\|\Phi\|_\alpha \,\eee^{-\alpha\card X}\, \eee^{\beta\|\Phi\|_\alpha}\;.
\end{equation}
We have used the trivial bound $e^x-1\le x\,\eee^x$ plus the fact that $\|\Phi(X)\| \le \|\Phi\|_\alpha \,\eee^{-\alpha\card X}$.
With this, convergence is guaranteed if
\begin{equation}\label{eq:greg.1}
\beta\,\|\Phi\|_\alpha\,\eee^{\beta\|\Phi\|_\alpha}\;\le\; \zeta(X)\, \exp\biggl[\biggl(\alpha -2\sup_y \sum_{\substack{Y\in\bonds\\ Y\ni y}} \zeta(Y)\biggr)\card X\biggr]\;.
\end{equation}
We must now choose $\zeta(X)$ so that it cancels the exponential dependence on $\card X$ in the right-hand side.  Here is a way.  Define
\begin{equation}\label{eq:greg.2}
\zeta(X)\;=\; \frac{\alpha}{4\,C_{\alpha/2}}\,\eee^{-(\alpha/2)\card X}
\end{equation}
with
\begin{equation}\label{eq:greg.3}
C_{\alpha/2}\;=\; \sup_y \sum_{\substack{Y\in\bonds\\ Y\ni y}} \eee^{-(\alpha/2)\card Y}
\end{equation}
chosen so that
\begin{equation}\label{eq:greg.4}
\sup_y \sum_{\substack{Y\in\bonds\\ Y\ni y}}\zeta(Y) \;=\; \frac{\alpha}{4}\;.
\end{equation}
With these choices, the right-hand side of \eqref{eq:greg.1} equals $\alpha/(4\,C_{\alpha/2})$ and we obtain
the following corollary of \ref{eq:cor1.bis}

\begin{cor}[General $\beta$-radius of convergence]\label{cor:gen-conv}
The cluster expansion converges for 
\begin{equation}\label{eq:gen-conv}
\beta\,\|\Phi\|_\alpha \,\eee^{\beta\|\Phi\|_\alpha} \;\le\; \frac{\alpha}{4\,C_{\alpha/2}}
\end{equation}
Furthermore, the bounds \eqref{eq:rr.2} and \eqref{eq:rr.2.bis} hold with the replacement
\begin{equation}
\eee^a\;=\; 1+ \frac{\alpha}{4}\;.
\end{equation}
\end{cor}
\begin{rem}
More generally, we can take any $\gamma\in(0,1)$ and consider 
\begin{equation}\label{eq:greg.6-conv}
\zeta(X)\;=\; \frac{(1-\gamma)\, \alpha}{2\,C_{(1-\gamma)\alpha}}\,\eee^{-(1-\gamma)\,\alpha\card X}
\end{equation}
to obtain
\begin{equation}\label{eq:gen-conv.1}
\beta\,\|\Phi\|_\alpha \,\eee^{\beta\|\Phi\|_\alpha}\ \;\le\; \frac{(1-\gamma)\,\alpha}{2\,C_{(1-\gamma)\alpha}}
\end{equation}
and
\begin{equation}
\eee^a\;=\; 1+ \frac{(1-\gamma)\,\alpha}{2}\;.
\end{equation}
\end{rem}
The radius \eqref{eq:gen-conv.1} can be optimized over $\gamma$ for each particular interaction.

\section{Results IV: Consequences of the convergence of the cluster expansion}\label{ssec:r.conseq.1}

We summarize the main physical consequences of a convergent cluster expansion.  We assume an interaction satisfying the exponential summability condition \eqref{eq:intro1} whose terms $\Phi(X)$ may depend analytically on a finite number of parameters.  The following results refer to two different parameter regions.  Free-energy analyticity holds in the region $\mathcal R_0$ of values of $\beta$ and additional parameters determined by condition \eqref{eq:rr.1} in Corollary \ref{eq:cor1}, which includes the smaller region determined by the more manageable sufficient condition \eqref{eq:min.6} in Corollary \ref{eq:cor1.bis}.   The convergence of the correlation expansion holds, instead, in a smaller region
$\widehat{\mathcal R}_0$ (see below)
.  In particular, the latter includes the 
``universal condition" \eqref{eq:gen-conv} [or \eqref{eq:gen-conv.1}] in Corollary \ref{cor:gen-conv}.
The difference between these regions could be a technical issue, due to the relatively unsophistication of the bounds used, and it is almost negligible in usual cases in which the weights $\zeta$ are small (see, e.g., the values in Table \ref{table:chi}).  Proofs are presented in Section \ref{sec:proofs}.   

\subsection{{\bf Free-energy density}}\label{ssec:r.free-energy}
The convergence of the free energy density requires two additional ingredients: translation invariance and convergence in the sense of van Hove.  Regarding the former, to avoid nonessential complications, it is enough to assume that $\lat=\mathbb Z^d$.  Expressions can be naturally extended, however, to periodic lattices, namely countable sets $\lat$ on which $\mathbb Z^d$ acts by homomorphisms and determines a finite elementary cell $\lat/\mathbb Z^d$.  In both cases, each point $x\in\mathbb Z^d$ defines translation operators acting on different objects but denoted equally $\tau_x$.  They are:
\begin{itemize}
\item[(i)] \emph{Translation of sets:} For each $A\subset\lat$,
\begin{equation}
\tau_x A \;=\; A+x\;=\; \bigl\{a+x: a\in A\bigr\}\;.
\end{equation}

\item[(ii)] \emph{Translation of finite spin configurations:} For each $\Lambda \sset\lat$ and $\omega_\Lambda\in\hsp_\Lambda$, there is a translated configuration $\tau_x\omega_\Lambda\in \hsp_{\Lambda+x}$ with 
\begin{equation}
\bigl(\tau_x \omega_\Lambda\bigr)_y \;=\; \omega_{y-x} \quad, y\in \Lambda+x\;.
\end{equation}

\item[(iii)] \emph{Translation of observables:} For each $\Lambda \sset\lat$ and local observable $F\in\alg_\Lambda$, there is a translated observable $\tau_x F\in \alg_{\Lambda+x}$ with 
\begin{equation}
\bigl(\tau_x F\bigr)(\omega_{\Lambda+x})\;=\; F(\omega_\Lambda) 
\end{equation}
that is, $\tau_x F=F\circ \tau_{-x}$.

\item[(iv)] \emph{Translation of states:} For each state $\Theta$ , its translate $\tau_x\Theta$ is defined by the identities 
\begin{equation}
\bigl(\tau_x \Theta \bigr)(F)\;=\; \Theta\bigl(\tau_{-x}F\bigr) \quad,\quad F\in \alg^{\rm local}
\end{equation}
that is, $\tau_x \Theta=\Theta\circ \tau_{-x}$.
\end{itemize}

\begin{defin}[Translation invariance]. For a quantum spin system, 
\begin{itemize}
\item[(a)] The sets of bonds is {translation invariant} if $X\in\bonds \;\longrightarrow\; \tau_x X\in \bonds$ for each $x\in\mathbb Z^d$.
\item[(b)] The interaction is translation invariant if so is the set of bonds and, furthermore, $\tau_x\Phi(X)=\Phi(\tau_{x} X)$, for each $x\in\mathbb Z^d$ and $X\in\bonds$.
 \item[(c)] A state $\Theta$ is translation invariant if $\tau_x\Theta=\Theta$ for each $x\in\mathbb Z^d$.
\end{itemize}

\end{defin}

\begin{defin}[van Hove families].  A family of finite sets $\{\Lambda\}\subset \lat$ is a \empbf{van Hove family} for $\lat$ if 
\begin{itemize}
\item[(i)] It is \empbf{exhaustive}:  Each finite subset of $\lat$ is contained in some $\Lambda$ of the family.  
\item[(ii)] Sets have \empbf{negligible area contribution}:  There is a distance ${\rm dist}$ in $\lat$ such that,
\begin{equation}
\frac{\card{\partial_\epsilon\Lambda}}{\card\Lambda}\,\longrightarrow\, 0 \quad\mbox{with}\quad 
\partial_\epsilon\Lambda \;=\; \bigl\{x\in\Lambda: {\rm dist}(x,\lat\setminus\Lambda)\le\epsilon\bigr\} \;
\end{equation}
and the limit is understood in the set-inclusion sense.  
\end{itemize}
  \end{defin}
Typically, van Hove sequences are chosen to be increasing sequences of simple geometric sets, such as squares or rectangles with sides growing at a comparable rate.  For the case $\lat=\mathbb Z^d$, or for periodic lattices, all usual unbounded distances are appropriate (e.g., Euclidean or $\ell_\infty$). 

\begin{cor}[{\bf Analyticity of the free energy}]\label{cor:conse0}
Consider a translation-invariant interaction and the resulting free energy densities
\begin{equation}
\beta f^{\#}_\Lambda\;=\; -\frac{1}{\card\Lambda}\, \log Z^{\#}_\Lambda\quad,\quad
\# = \Theta, \mbox{per}, \emptyset
\end{equation}
and the thermodynamic limits (set-inclusion sense)
\begin{equation}\label{eq:rr.;imf}
 f^\#\;=\; \lim_{\Lambda\to\mathbb{L}}  f^\#_\Lambda  \;.
\end{equation}
Then, uniformly within the region $\mathcal R_0$ defined by condition \eqref{eq:rr.1} 
\begin{itemize}
\item[(i)] The limits \eqref{eq:rr.;imf} exist for each van Hove family $\{\Lambda\}$ and coincide with the density $f$ for free boundary conditions, which, furthermore, takes the form
\begin{equation}\label{eq:tan.1}
\beta f \;=\;-\sum_{\substack{\mathcal C\in \mathbb C_{\B_{\lat}}\\ \underline{\mathcal C}\ni 0}} \frac{\Omega^T(\mathcal C)}{\card{\underline{\mathcal C}}}
\end{equation}
with
\begin{equation}
\Omega^{T} (\B_1,\ldots,\B_n)\;=\; \frac{1}{n!} \,\omega_n^T(\B_1,\ldots,\B_n) \,\rho_{\B_1}\ldots \rho_{\B_{n}}\;.
\end{equation}
\item[(ii)] The thermodynamic free energy $\beta f$ and all the finite-volume densities $\beta f^\#_\Lambda$ are analytic functions of $\beta$ and of any finite family of parameters on which the interaction terms have an analytic dependence.
\item[(iii)]  Free energy densities obey the bounds
\begin{equation}\label{eq:tan.2}
\bigl|\beta f^\#_\Lambda\bigr|\,, \card{\beta f} 
\;\le\; \eee^a-1\;.
\end{equation}
\end{itemize}
\end{cor}

\begin{rem} Regarding item (a), let us point out that both the existence of the limits \eqref{eq:rr.;imf} and their insensitivity to boundary conditions are valid in general throughout the phase diagram.  The convergence of the cluster expansion leads, in addition, to the explicit expression \eqref{eq:tan.1} and the bounds \eqref{eq:tan.2}.  
\end{rem}

\subsection{{\bf Correlations and KMS states}}

\begin{cor}[{\bf Analyticity of correlations}] \label{eq:conse4}
Consider the reduced correlations
\begin{equation}\label{eq:conse5}
g_{\Lambda}^{\#}(X_0):=\frac{Z^{\#}_{\Lambda\setminus X_0}}{Z^{\#}_{\Lambda}}\quad,\quad
\# = \Theta, \mbox{per}, \emptyset
\end{equation}
for finite $X_0\subset\Lambda$.  Then, uniformly within the region $\mathcal R_0$, the limits
\begin{equation}
g^\#(X_0)\;:=\;\lim_{\Lambda\uparrow\mathbb L}g^\#_{\Lambda}(X_0)
\end{equation}
exist, coincide, are analytic in $\beta$ and additional interaction parameters, and their common value is the expansion
\begin{equation}\label{eq:tan.1}
g(X_0) \;=\;\exp\Biggl[\sum_{\substack{\mathcal C\in \mathbb C_{\B_{\lat}}\\ \underline{\mathcal C}\cap X_0\neq \emptyset}} \Omega^T(\mathcal C)\Biggr]\;.
\end{equation}
Furthermore,
\begin{equation}\label{eq:rr.eli.20}
\bigl |g^\#_\Lambda(X_0)\bigr|\,, \card{g(X_0)}\;\le\; \exp\bigl[\card{X_0}(\eee^a-1)\bigr]
\;=\; \; \exp\biggl[\card{X_0}\sup_y \sum_{\substack{Y\in\bonds\\ Y\ni y}} \zeta(Y)\biggr]\;.
\end{equation}
\end{cor}

The following corollary refers to the region $\widehat{\mathcal R}_0$ defined by the condition
\begin{equation}\label{eq:min.6.exp}
\eee^{\beta\|\Phi(X)\|}-1\;\le\; \frac{\zeta(X)}{\displaystyle \exp\biggl[\card X\sup_y \sum_{\substack{Y\in\bonds\\ Y\ni y}} \zeta(Y)\biggr]
\prod_{\substack{Y\in\bonds \\ Y\cap X\neq\emptyset\\Y\neq X}} \bigl[1+\zeta(Y)\bigr]}
\end{equation}
for some function $\zeta:\bonds \to (0,\infty)$.  Note that this region includes the one defined by condition \eqref{eq:min.6.b} and, thus, by the ``universal" radius \eqref{eq:gen-conv} and weights \eqref{eq:greg.2} [or condition \eqref{eq:gen-conv.1} and weights \eqref{eq:greg.6-conv}].

\begin{cor}[{\bf Existence of Gibbs (=KMS) states}]\label{eq:conse2} Uniformly within the region $\widehat{\mathcal R}_0$, for each local observable $F$ with support $X_0$, the limits
\begin{equation}\label{eq:conse3.L}
\pi^{\#}(F)=\lim_{\Lambda\uparrow\mathbb L} \pi_\Lambda(F)\quad,\quad
\# = \emptyset\,, \Theta\,, \mbox{per}
\end{equation}
exist, coincide, are analytic in $\beta$ and other interaction parameters and their common value is given by the expansion 
\begin{equation}\label{eq:pi.exp}
\pi(F)\;=\; \sum_{\substack{\B\subset\bonds\\ \B\cup X_0\,{\rm connected}\\ \B\not\sim X_0}} K(A,\B) \,g (X_0\cup\underline\B)
\end{equation}
with
\begin{equation}\label{eq:quan8.ave.20.L}
 K(F,\B)  \;:=\; \sum_{\widetilde\B\subset\B}
(-1)^{\card{\B\setminus\widetilde\B}} \,
\tr_{X_0\cup\underline\B} \Big( A\,\eee^{-\beta H_{\widetilde\B}}\Bigr)
\end{equation}
and
\begin{equation}\label{eq:g.L}
g(X_0\cup\underline\B)\;=\; \lim_{\Lambda\uparrow\mathbb L} \frac{Z_{\Lambda\setminus (X_0\cup\underline\B)}}{Z_\Lambda} \;=\; \exp\Biggl[\sum_{\substack{\mathcal C\in \mathbb C_{\B_{\lat}}\\ \underline{\mathcal C}\cap( X_0\cup\underline\B)\neq\emptyset}} \Omega^T(\mathcal C)\Biggr]\;.
\end{equation}
Furthermore, 
\begin{equation}\label{eq:g.M}
\bigl| \pi_\Lambda^{\#}(F) \bigr|\,, \bigl| \pi^{\#}(F) \bigr| \;\le\; \|F\|\,
\exp\biggl[\card{X_0}\sup_y \sum_{\substack{Y\in\bonds\\ Y\ni y}} \zeta(Y)\biggr]
\prod_{\substack{Y\in\bonds \\ Y\cap X_0\neq\emptyset}} \bigl[1+\zeta(Y)\bigr]\;.
\end{equation}

\end{cor}.
\begin{rems}
\begin{description}

\item[(R1)] By general KMS theory (see, e.g.~Proposition 6.2.15 in \cite{brarob2}), the linear form $\pi$ is a KMS state. 
\item[(R2)] Compactness arguments imply the existence of the limits \eqref{eq:conse3.L} through subsequences of volumes.  In the region $\mathcal R_0$ the limit exist, in fact, in the strong set-inclusion set and without having to pass to  subsequences. 
\end{description} 
\end{rems}

\section{Proofs}\label{sec:proofs}


\subsection{{\bf Proofs for Sections \ref{ssec:rr.tan.10} and  \ref{ssec:rr.tan.10.int} }}

\subsubsection{{\bf Proof of Theorem \ref{theo:rr.tan.06} and Corollary \ref{cor:bio.1}}} 
Theorem \ref{theo:rr.tan.06}  is an abridged version of a well known result \cite{PF07,FN19}.  For completeness we review its full statement and its relatively straightforward proof in Appendix \ref{app:sketch}.
The absolute convergence of the free-energy expansion \eqref{eq:polmect10.1} follows from identity \eqref{eq:rr.bul.1}.  To prove inequality \eqref{eq:rr.bul.3} in Corollary \ref{cor:bio.1} we combine \eqref{eq:rr.bul.1} and \eqref{eq:rr.bul.1.1}: 
\begin{equation}\label{eq:rr.eli.10}
\frac{Z_{\B_\Lambda\setminus \B_0}}{Z_{\Lambda}}\;=\; \frac{\partial}{\partial\rho_{\B_0}} \ln Z_{\Lambda}(\bbd\rho)\;=\; \Sigma^\Lambda_{\B_0}(\bbd{\rho})\;.
\end{equation}
Hence, if $\card{\bbd\rho} \le \bbd \lambda$, with $\bbd\lambda$ and $\bbd\mu$ satisfying \eqref{eq:quan23}, 
\begin{equation}\label{eq:rr.oeg.202}
\left|\frac{Z_{\B_\Lambda\setminus \B_0}}{Z_{\Lambda}}\right|\;\le\;
\card{\Sigma}^\Lambda_{\B_0}(\bbd{\lambda})\;\le\; \card{\Sigma}_{\B_0}(\bbd{\lambda})
\;\le\; \frac{\mu(\B_0)}{\card{\rho(\B_0)}}\;.
\end{equation}
The last inequality is due to inequality in \eqref{eq:mmm.1}.
We observe that this bound can be improved using previous bounds in the chain of inequalities \eqref{eq:polmect10.1.3}. 
Furthermore, inequalities \eqref{eq:rr.bul.2} together with the criterion \eqref{eq:quan23} yield the weaker inequalities
\begin{equation}\label{eq:rr.bul.2.bis}
\card{\log \frac{ Z_\Lambda}{ Z_{\B_\Lambda\setminus \B}}} \;\le\;
-\varphi_\B(\bbd\mu)\,\log\Bigl(1-\frac{\mu_\B}{\varphi_\B(\bbd\mu)}\Bigr)\;\le\; \log\bigl(1+\mu_\B\bigr)\;\le\; \mu_\B\;.
\end{equation}

%
%
%
%

\subsubsection{{\bf Proof of Lemma \ref{Lemma2}}}
It is enough to prove \eqref{eq:rr.oeg.trad}, as inequality \eqref{eq:rr.bound} follows immediately from the fact that the passage from $\xi$ to $\rho$ involves a normalized trace, plus the fact that, by simple observation,
\begin{equation}
\| \Phi^\Theta(X)\| \,,\, \| \Phi^{\rm per}(X)\| \;\le\; \| \Phi(X)\| 
\end{equation}

In the sequel we spell out a proof of \eqref{eq:rr.oeg.trad} presented by Park \cite{Park82}.  The starting expression, given in the next proposition, is an instance of the use of decoupling parameters discussed in Appendix \ref{ssec:decoup}.  We introduce parameters $\underline s=\{s_X: X\in\bonds\}\in[0,1]^{\bonds}$ and fix some finite $\B_\Lambda\subset\bonds$.  For disjoint finite families of bonds $\B'$ and $\B$ in $\B_\Lambda$ we define
\begin{equation}
H_{\B'}(\underline s_{\B})\;=\; \sum_{X\in\B'} \Phi(X) + \sum_{X\in\B} s_X\,\Phi(X)
\end{equation}
and
\begin{equation}
\xi_\B(\underline s_{\B_{\Lambda}\setminus\B})\;=\;  \sum_{\B'\subset\B} (-1)^{\card{\B\setminus\B'}} \, \eee^{-\beta\,H_{\B'}(\underline s_{\B_\Lambda\setminus\B})}
\end{equation}
We remark that
\begin{equation}
H_\emptyset(\underline s_\Lambda)\;=\; \sum_{X\in\B_\Lambda} s_X\,\Phi(X) 
\quad\mbox{and}\quad \xi_{\B_\Lambda}\;=\; \xi_{\B_\Lambda}(\underline s_\emptyset)\;.
\end{equation}

The key expression is the following.
\begin{prop}
For any $\B\subset\B_\Lambda$, 
\begin{equation}\label{eq:opera.1}
\xi_{\B}(\underline s_{\B\Lambda\setminus\B}) \;=\; \prod_{X\in\B}\Bigl[\int_0^1 ds_X\,\frac{\partial\;}{\partial s_X}\Bigr] \eee^{-\beta\,H_{\emptyset}(\underline s_{\B\Lambda\setminus\B})}
\end{equation}
if $\B\neq\emptyset$, and $\xi_\emptyset(\underline s)= \eee^{-\beta\,H_{\emptyset}(\underline s)}$.
\end{prop}
\proof The expression follows by induction on $\card\B$.  For $\card\B=0$ the formula is true by definition.  The inductive step is as follows:
 Classifying families $\B'$ according on whether they contain a fixed $X\in\bonds$ or not,  we obtain .
\begin{eqnarray}
\lefteqn{\xi_{\B\cup\{X\}}(\underline s_{\B_\Lambda\setminus(\B\cup\{X\}})
 \;=\; \sum_{\B'\subset\B\cup\{X\}} (-1)^{\card{\B\cup\{X\}\setminus\B'}}\,\eee^{-\beta H_{\B'}(\underline s_{\B_\Lambda\setminus(\B\cup\{X\})})} }\nonumber\\
&=& \sum_{\B'\subset\B} (-1)^{\card{\B\cup\{X\}\setminus\B'}}\,\eee^{-\beta H_{\B'}(\underline s_{\B_\Lambda\setminus(\B\cup\{X\})})} +
\sum_{\B'\subset\B} (-1)^{\card{\B\cup\{X\}\setminus\B'\cup\{X\}}}\,\eee^{-\beta H_{\B'\cup \{X\}}(\underline s_{\B_\Lambda\setminus(\B\cup\{X\})})} \nonumber\\
&=& \sum_{\B'\subset\B} (-1)^{\card{\B\setminus\B'}} \Bigl[ \eee^{-\beta H_{\B'\cup \{X\}}(\underline s_{\B_\Lambda\setminus(\B\cup\{X\})})} - \eee^{-\beta H_{\B'}(\underline s_{\B_\Lambda\setminus(\B\cup\{X\})})}\Bigr]\;.
\end{eqnarray}

Hence,
\begin{eqnarray}
\xi_{\B\cup\{X\}}(\underline s_{\B\Lambda\setminus(\B\cup\{X\}}) &=& \sum_{\B'\subset\B} (-1)^{\card{\B\setminus\B'}} \int_0^1 ds_X\,\frac{\partial\;}{\partial s_X}
\eee^{-\beta H_{\B'}(\underline s_{\B_\Lambda\setminus\B})}\nonumber\\
&=& \int_0^1 ds_X\,\frac{\partial\;}{\partial s_X} \xi_\B(\underline s_{\B_\Lambda\setminus\B})\;.
\end{eqnarray}
To prove \eqref{eq:opera.1} we apply the inductive step for some fixed order of the bonds in $\B$.  Expression \eqref{eq:opera.1} holds whatever the order.  \qed 

As \eqref{eq:opera.1} holds for every $\B_\Lambda\supset\B$, in particular we can take $\B_\Lambda=\B$ to obtain
\begin{equation}\label{eq:opera.2}
\xi_{\B}\;=\; \prod_{X\in\B}\Bigl[\int_0^1 ds_X\,\frac{\partial\;}{\partial s_X}\Bigr] \eee^{-\beta\,H_{\emptyset}(\underline s_{\B})}
\end{equation}

To bound \eqref{eq:opera.2} we iteratively use the well known identity given in the following lemma.
\begin{lem}
If $B(s)$ is a family of bounded operators with a differentiable dependence in $s$,
\begin{equation}\label{eq:opera.3}
\frac{d}{d s} \eee^{B(s)}\;=\; \int_0^1 dt\,\eee^{t B(s)}\,\Bigl[\frac{d}{ds} B(s)\Bigr]\,\eee^{(1-t) B(s)}\;.
\end{equation}
\end{lem}
\proof The formula is an immediate consequence of the identity
\begin{equation}\label{eq:intvir}
\eee^D - \eee^C\;=\; \int_0^1 dt\;\eee^{tD} (D-C) \,\eee^{(1-t)}
\end{equation}
valid for all finite-dimensional operators $D,C$.  This identity follows from the fact that both sides are equal to
\begin{equation}
\int_0^1 dt\, \frac{d\;}{ds}\Bigl[\eee^{tD}\,\eee^{(1-s)B}\Bigr]\;.
\end{equation}
To obtain \eqref{eq:opera.3} we apply \eqref{eq:intvir} to $D=B(s+h)$, $C=B(s)$, divide both sides by $h$ and take the limit $h\to 0$. \qed

In the case of interest here, the operator $H_\empty(\underline s)$ has a linear dependence on each parameter $s_X$.  From \eqref{eq:opera.3} we get
\begin{equation}\label{eq:opera.4}
\frac{\partial\;}{\partial s_X} \eee^{-\beta\,H_{\emptyset}(\underline s_{\B})}\;=\; \int_0^1 dt_X\,
\eee^{-t_X\beta\,H_{\emptyset}(\underline s_{\B})} \bigl[-\beta\Phi(X)\bigr]
\eee^{-(1-t_X)\beta\,H_{\emptyset}(\underline s_{\B})}\;,
\end{equation}
which yields the bound
\begin{equation}
\Bigl\| \frac{\partial\;}{\partial s_X} \,\eee^{-\beta\,H_{\emptyset}(\underline s_{\B})}\Bigr\|\;\le\; \beta\|\Phi(X)\|\,
\eee^{\beta\,\|H_{\emptyset}(\underline s_{\B}\|}\;. 
\end{equation}

Application of further derivatives yield progressively more complicated expressions that, however, are relatively easy to bound.  To see the pattern, let us apply a further derivative to \eqref{eq:opera.4}. By Leibnitz rule, this yields the sum of two terms, each involving the derivative of one of the exponentials in the right-hand side of \eqref{eq:opera.4}.  Applying the latter to each of these terms we finally obtain

\begin{eqnarray}
\frac{\partial\;}{\partial s_Y} \frac{\partial\;}{\partial s_X} \eee^{-\beta\,H_{\emptyset}(\underline s_{\B})} &=&
\int_0^1 dt_X\, \frac{\partial\;}{\partial s_Y}
\eee^{-t_X\beta\,H_{\emptyset}(\underline s_{\B})} \bigl[-\beta\Phi(X)\bigr]
\eee^{-(1-t_X)\beta\,H_{\emptyset}(\underline s_{\B})}\nonumber\\
&&{}+  \int_0^1 dt_X\,
\eee^{-t_X\beta\,H_{\emptyset}(\underline s_{\B})} \bigl[-\beta\Phi(X)\bigr]
\,\frac{\partial\;}{\partial s_Y}\eee^{-(1-t_X)\beta\,H_{\emptyset}(\underline s_{\B})}\;.
\end{eqnarray}
Applying \eqref{eq:opera.4} for each derivative we conclude that
\begin{eqnarray}
\lefteqn{\frac{\partial\;}{\partial s_Y} \frac{\partial\;}{\partial s_X} \eee^{-\beta\,H_{\emptyset}(\underline s_{\B})}\;=\;}\nonumber\\
&&\int_0^1 dt_X\, \int_0^1 dt_Y\Bigl\{\eee^{-t_Yt_X\beta\,H_{\emptyset}(\underline s_{\B})} \bigl[-t_X\beta\Phi(Y)\bigr]
\eee^{-(1-t_Y)t_X\beta\,H_{\emptyset}(\underline s_{\B})} \bigl[-\beta\Phi(X)\bigr]
\eee^{-(1-t_X)\beta\,H_{\emptyset}(\underline s_{\B})}\nonumber\\
&&{}+ \eee^{-t_X\beta\,H_{\emptyset}(\underline s_{\B})} \bigl[-\beta\Phi(X)\bigr]
\eee^{-t_Y(1-t_X)\beta\,H_{\emptyset}(\underline s_{\B})} \bigl[-(1-t_X)\beta\Phi(Y)\bigr]
\eee^{-((1-t_Y)1-t_X)\beta\,H_{\emptyset}(\underline s_{\B})}\Bigr\}\;.
\end{eqnarray}
As the coefficients of $\beta H_\emptyset$ in each term add up to one, and so do the coefficients of $\beta\Phi(Y)$, we get the bound
\begin{equation}\label{eq:opera.6}
\Bigl\|\frac{\partial\;}{\partial s_Y} \frac{\partial\;}{\partial s_X} \eee^{-\beta\,H_{\emptyset}(\underline s_{\B})}\Bigr\|\;\le\;
\beta\|\Phi(X)\|\, \beta\|\Phi(Y)\|\,
\eee^{\beta\,\|H_{\emptyset}(\underline s_{\B}\|}\;. 
\end{equation}

The final expression is as follows.
\begin{prop}
Assume $\B=\{X_1,\ldots,X_n\}$.  Then,
\begin{eqnarray}
\lefteqn{\prod_{X\in\B}\frac{\partial\;}{\partial s_X} \eee^{-\beta\,H_{\emptyset}(\underline s_{\B})}\;=\;}\nonumber\\
&&\sum_{\pi \; n-{\rm permut.} }\eee^{-\underline{\widetilde t}_{\pi(1)} \beta\,H_{\emptyset}(\underline s_{\B})}
\bigl[-\underline{\widehat t}_{\pi(1)}\beta \Phi(X_{\pi(1)}\bigr]\,\cdots\, 
\eee^{-\underline{\widetilde t}_{\pi(n)} \beta\,H_{\emptyset}(\underline s_{\B})}
\bigl[-\underline{\widehat t}_{\pi(n)}\beta \Phi(X_{\pi(n)}\bigr]\,
\eee^{-\bigl(1-\sum_{i=1}^n \underline{\widetilde t}_{\pi(i)}\bigr) \beta\,H_{\emptyset}(\underline s_{\B})}\nonumber\\
\ 
\end{eqnarray}
with
\begin{equation}
\underline{\widetilde t}_{\pi(i)}\;=\; \widetilde t^{\pi(i)}_1\cdot \,\cdots\, \widetilde t^{\pi(i)}_n \quad \mbox{with} \quad
\widetilde t^{\pi(i)}_j\in\{1,t_{X_j}, 1-t_{X_j}\}
\end{equation}
and 
\begin{equation}
\underline{\widehat t}_{\pi(i)}\;=\; \widehat t^{\pi(i)}_1\cdot \,\cdots\, \widehat t^{\pi(i)}_n \quad \mbox{with} \quad
\widehat t^{\pi(i)}_j\in\{1,t_{X_j}, 1-t_{X_j}\}
\end{equation}
such that
\begin{equation}
\sum_{\pi \; n-{\rm permut.} }\sum_{i,j=1}^n {\widehat t}^{\pi(i)}_j\;=\; 1\;.
\end{equation}
\end{prop}
The proposition is proven by straightforward induction on the number of derivatives.  We count with the reader's understanding if we omit the non-revealing and notationally involved formal proof.  The proposition immediately yields the generalization of \eqref{eq:opera.6}.
\begin{cor}
For any finite $\B\subset\bonds$,
\begin{equation}\label{eq:opera.7}
\Bigl\|\prod_{X\in\B}\frac{\partial\;}{\partial s_X} \eee^{-\beta\,H_{\emptyset}(\underline s_{\B})}\Bigr\| \;\le\;
\Bigl[\prod_{X\in\B} \bigl[\beta\bigl\|\Phi(X)\bigr\| \Bigr]\eee^{\beta\,\|H_{\emptyset}(\underline s_{\B})\|}\;.
\end{equation}
\end{cor}

Replacing \eqref{eq:opera.7} in \eqref{eq:opera.1}. we finally get
\begin{eqnarray}
\bigl\|\xi_\B\bigr\| &\le& \prod_{X\in\B} \int_0^1 ds_X \,\beta\|\Phi(X)\|\,\eee^{s_x\beta\|\Phi(X)\|} \nonumber\\[8pt]
&=& \prod_{X\in\B} \bigl[\eee^{\beta\|\Phi(X)\|} -1\bigr]\;,
\end{eqnarray}
proving Lemma \ref{Lemma2}.  \qed

\subsubsection{{\bf Proof of Corollary \ref{eq:cor1}}}
\begin{itemize}
\item[(a)] Criterion \eqref{eq:rr.1} and bound  \eqref{eq:polmect10.1.1.A} are a consequence of condition \eqref{eq:rr.convergence.1} with the choice \eqref{eq:choice}.
\item[(b)] Bounds \eqref{eq:rr.2} and \eqref{eq:rr.2.bis} are proven exactly as in the proof of Corollary \ref{cor:bio.1} above. 
\item[(c)] The bound \eqref{eq:site-pinned.b} is proven through the following inequalities:
\begin{equation}
\bigl|\widehat\Sigma\bigr|_{x}(\bbd{V})\;\le\; \sum_{\B\ni x} V_\B \,\bigl|\widehat\Sigma\bigr|_{\B}(\bbd{V})\;\le\;
\sum_{\B\ni x} V_\B \,\eee^{a\card{\B}}\;\le\; \eee^a-1\;.
\end{equation}
The second inequality is due to \eqref {eq:rr.convergence.1.bis} and the last one to 
\eqref{eq:rr.convergence.1}. \qed
\end{itemize}

\subsection{{\bf Proofs of Section \ref{ssec:r.conseq.1}}}\label{ssec:rr.proof.conseq}

\subsubsection{{\bf Proof of Corollary \ref{cor:conse0}}}
We fix an order of the sites in $\lat$ and consider, for each finite $\Lambda\sset\lat$ and each $x\in\lat$,
\begin{equation}\label{eq:polmect10.1.1.A}
h^\#_\Lambda(x)\;=\; \sum_{\substack{\mathcal C\in \mathbb C_{\B^\#_\Lambda}\\\underline{\mathcal C} \ge x}} \Omega^{T,\#}(\mathcal C)\;,
\end{equation}
where, for $A\subset\lat$, the notation $A\ge x$ means that $x$ is the first point in $A$ according to the fixed order.

\begin{lem}\label{lem:pinned-x}
\begin{equation}\label{eq:polmect10.1.1.A}
\lim_{\Lambda\uparrow\lat} h^\#_\Lambda(x)\;=\; h(x)\;:=\;\sum_{\substack{\mathcal C\in \mathbb C_{\B_\lat}\\\underline{\mathcal C} \ge x}} \Omega^{T}(\mathcal C)\;.
\end{equation}
\end{lem}
\proof
\begin{equation}
\bigl| h(x)- h^\#_\Lambda(x)\bigr| \;\le\; 2 \sum_{n=1}^{\infty}\frac{1}{n!}\sum_{\substack{(\B_1,\ldots,\B_n)\in\pro^n\\
\cup_i\underline{\B}_i\ge x\\
\bigl(\cup_i\underline{\B}_i\bigr)\cap\bigl(\lat\setminus\Lambda\bigr)\neq\emptyset}}\card{\omega_n^T(\B_0,\B_1,\ldots,\B_n)}\,V_{\B_1}\cdots V_{\B_{n}}
\end{equation}
which converges to zero, in set-inclusion sense, due to the convergence of 
$\bigl|\widehat\Sigma\bigr|_{x}(\bbd{V})$\;. \qed
\smallskip

\paragraph{\bf Proof of (i)} 
We will prove that
\begin{equation}\label{eq:free.great}
\lim_{\Lambda\uparrow\lat} -\beta f^\#_\Lambda \;=\; h(0)\;=\; 
\sum_{\substack{\mathcal C\in \mathbb C_{\B^\#_\Lambda}\\\underline{\mathcal C} \ge 0}} \Omega^{T,\#}(\mathcal C)\;.
\end{equation}
Formula \eqref{eq:tan.1} follows form the fact that, by translation invariance,
\begin{equation}\label{eq:free.great}
 \Omega^{T}(\mathcal C)\;=\; \frac{1}{\card{\underline{\mathcal C}}} \sum_{x\in \underline{\mathcal C}} \Omega^{T}(\tau_x\mathcal C);.
\end{equation}
The expression \eqref{eq:tan.1} is preferred over the totally equivalent expression \eqref{eq:free.great} because the former is uniquely defined while the latter makes use of a reader-dependent arbitrary order of the sites. 
The analyticity follows from the fact that the series of $h^\#(0)$ and $h(0)$ ---and hence those of $\beta f^\#_\Lambda$ and $\beta f$--- are termwisely dominated by the series $\bigl|\widehat\Sigma\bigr|_{x}(\bbd{V})$.

The proof of \eqref{eq:free.great} relies on the following ingredients.
\begin{itemize}
\item[(I1)] The cluster-expansion identity
\begin{equation}
\log Z^\#_\Lambda \;=\; \sum_{x\in\Lambda} h^\#_\Lambda(x)\;.
\end{equation}

\item[(I2)] The fact that, by the previous lemma, for each given $\varepsilon>0$ there exists a region $\Gamma_\varepsilon$ such that
\begin{equation}
\bigl| h(0)- h^\#_{\Gamma}(0)\bigr| \;\le\; \varepsilon
\end{equation}
for each $\Gamma\supset\Gamma_\varepsilon$
and, therefore, by translation invariance,
\begin{equation}\label{eq:hxepsilon}
\bigl| h(x)- h^\#_{\Gamma}(x)\bigr| \;\le\; \varepsilon
\end{equation}
for every $x\in\lat$ and $\Gamma\supset\Gamma_\varepsilon$.  

\item[(I3)] The lattice-wise bound
\begin{equation}\label{eq:hxbound}
\bigl| h^\#_{\Lambda}(x)\bigr| \;\le\; \bigl|\widehat\Sigma\bigr|_{x}(\bbd{V})\;\le\; \eee^a-1
\end{equation}
which follows from \eqref{eq:site-pinned.b}. 

\item[(I4)] As a consequence of \eqref{eq:hxepsilon}, each finite $\Lambda\subset\lat$ contains the set
\begin{equation}
\Lambda_0 \;=\; \bigl\{ x\in\Lambda: \Gamma_\varepsilon +x \subset \Lambda\bigr\}
\end{equation}
which is such that 
\begin{equation}\label{eq:hxepsilon.bound}
\bigl| h(x)- h^\#_{\Lambda}(x)\bigr| \;\le\; \varepsilon
\end{equation}
for each $x\in\Lambda_0$.

\item[(I5)] If $\delta={\rm diam}{\Gamma_\varepsilon}$,
\begin{equation}
\Lambda\setminus\Lambda_0\;\subset\; \partial_\delta \Lambda
\end{equation}
\end{itemize}

From (I1) and translation invariant we obtain
\begin{equation}
-\beta f^\#_\Lambda - h(0)\;=\; \frac{1}{\card\Lambda} \sum_{x\in\Lambda} \bigl[h^\#_\Lambda(x)-h^\#(x)\bigr]\;.
\end{equation}
Splitting the sum into sites inside and outside $\Lambda_0$ yields
\begin{eqnarray}
\bigl|-\beta f^\#_\Lambda - h(0)\bigr| &\le& 
 \frac{1}{\card\Lambda} \sum_{x\in\Lambda_0} \bigl|h^\#_\Lambda(x)-h^\#(x)\bigr|
 +  \frac{1}{\card\Lambda} \sum_{x\in\Lambda\setminus\Lambda_0} \bigl|h^\#_\Lambda(x)-h^\#(x)\bigr|\nonumber\\
 &\le& \frac{\card{\Lambda_0}}{\card\Lambda} \varepsilon +  \frac{\card{\Lambda\setminus\Lambda_0}}{\card\Lambda} \bigl(\eee^a-1\bigr)\;.
\end{eqnarray}
The last line is due to the bounds given in (I2) and (I3).  Finally, due to (I5) plus the fact that the sets $\Lambda$ form a van Hove sequence, we conclude
\begin{equation}
\lim_{\Lambda\uparrow\lat} \bigl|-\beta f^\#_\Lambda - h(0)\bigr| \;\le\; \varepsilon
\end{equation}
for arbitrary $\varepsilon$.  \qed
\smallskip

\paragraph{\bf Proof of (ii)}
From the leftmost inequality in \eqref{eq:hxbound} 
\begin{equation}\label{eq:flgound}
\bigl|\beta f^\#_\Lambda\bigr| \;\le\; \frac{1}{\card\Lambda} \sum_{x\in\Lambda}\bigl|\widehat\Sigma\bigr|_{x}(\bbd{V})
\end{equation}
where, in fact, the bound holds termwisely for the corresponding power series.  Hence each $f^\#_\Lambda$ inherits the analyticity of the positive series $\bigl|\widehat\Sigma\bigr|_{x}(\bbd{V})$.  Passing to limits (Lemma \ref{lem:pinned-x}), the leftmost inequality \eqref{eq:hxbound} (plus translation invariance) implies
\begin{equation}\label{eq:h0bound}
\bigl|\beta f \bigr| \;=\;\bigl| h(0)\bigr| \;\le\; \bigl|\widehat\Sigma\bigr|_{0}(\bbd{V})
\end{equation}
also in the sense of termwise majorization of power series.  This proves the analyticity of $f$.  \qed
 \smallskip

\paragraph{\bf Proof of (iii)} This follows from the bounds \eqref{eq:flgound}, \eqref{eq:h0bound} and   \eqref{eq:site-pinned.b} [rightmost inequality in \eqref{eq:hxbound}].  \qed

\subsubsection{{\bf Proof of Corollary \ref{eq:conse4}}}
Writing the partition functions as the exponentials of the cluster expansions \eqref{eq:polmect10.1} (in the ``$\#$" version) and canceling out common terms in both exponentials we see that, within the convergence region, 
\begin{equation}\label{eq:tantan.1}
g^\#_\Lambda(X_0) \;=\;\exp\Biggl[\sum_{\substack{\mathcal C\in \mathbb C_{\B_{\Lambda}}\\ \underline{\mathcal C}\cap X_0\neq \emptyset}} \Omega^T(\mathcal C)\Biggr]\;.
\end{equation}
The fact that 
\begin{equation}
\lim_{\Lambda\uparrow\lat} g^\#_\Lambda(X_0) \;=\; g(X_0) 
\end{equation}
follows from the fact that 
\begin{equation}
\lim_{\Lambda\uparrow\lat} \sum_{\substack{\mathcal C\in \mathbb C_{\B_{\Lambda}}\\ \underline{\mathcal C}\cap X_0\neq \emptyset}} \Omega^T(\mathcal C)\;=\;
\sum_{\substack{\mathcal C\in \mathbb C_{\B_{\lat}}\\ \underline{\mathcal C}\cap X_0\neq \emptyset}} \Omega^T(\mathcal C)\;.
\end{equation}
This convergence is, in turns, a consequence of the fact that the difference between left- and right-hand terms is bounded above, in absolute value, by the tail of $\sum_{x\in X_0}\bigl|\widehat \Sigma\bigr|_{x}(\bbd V)$, which converges to zero as $\Lambda$ grows due to the summability of the latter.  The analyticity of each $g^\#_\Lambda(X_0)$ and $ g(X_0)$ is a consequence of the bounds 
\begin{equation}\label{eq:rr.eli.20.proof}
\bigl |g^\#_\Lambda(X_0)\bigr|\,, \card{g(X_0)}\;\le\; \exp\Bigl[\sum_{x\in X_0}\bigl| h^\#_{\Lambda}(x)\bigr|\Bigr]\;\le\;
\exp\Bigl[\sum_{x\in X_0}\bigl|\widehat \Sigma\bigr|_{x}(\bbd V)\Bigr]\;.
\end{equation}
The last inequality together with the bound \eqref{eq:site-pinned.b} yield \eqref{eq:rr.eli.20}.  The last identity in this formula is due to the choice \eqref{eq:min.5}. \qed
  
We remark that Park \cite{Park82} had to resort to the Peierls-Bogoliubov inequality to estimate these reduced correlations function.  Our expansions makes the estimation more self-contained and precise. 

\subsubsection{{\bf Proof of Corollary \ref{eq:conse2}}}

It is enough to uniformly bound $\pi^\#_\Lambda(F)$ and $\pi(F)$\eqref{eq:conse3.L} by a convergent positive power series.  As the difference $\bigl|\pi^\#_\Lambda(F) - \pi(F)\bigr|$ is bounded by the tail of the series, the summability of the latter implies convergence as $\Lambda\uparrow \lat$.  Analiticity is also a consequence of the termwise absolute uniform bound by a convergent power series.  The bounding series is obtained by combining the following two bounds.  The first bound is a consequence of \eqref{eq:rr.bound},
\begin{equation}
\bigl| K^\#_\Lambda(F,\B)\bigr|\,, \bigl| K^\#_\Lambda(F,\B)\bigr|\le\; \|F\| \,\|\rho^\#(\B)\|\;\le\; \|F\|\,
\prod_{X\in\B}\left(\eee^{\beta\|\Phi(X)\|}-1\right)\;.
\end{equation}
The second inequality is an application of the correlation inequality \eqref{eq:rr.eli.20} of the precedent Corollary \ref{eq:conse4}:
\begin{equation}\label{eq:rr.eli.20.horror}
\bigl |g^\#_\Lambda(X_0\cup \underline\B)\bigr|\,, \card{g(X_0\cup \underline\B)}\;\le\; \exp\bigl[\card{X_0\cup \underline\B}(\eee^a-1)\bigr]\;.
\end{equation}
These inequalities lead to
\begin{equation}\label{eq:opera.2}
\bigl|\pi^\#_\Lambda\bigr|\,, \bigl|\pi^\#_\Lambda\bigr|\;\le\; \|F\|\, \exp\bigl[\card{X_0}(\eee^a-1)\bigr]\,\widetilde T(X_0)
\end{equation}
with $\widetilde T(X_0)$ is the tree expansion \eqref{eq:every.2} but with weights
 \begin{equation}\label{eq:min.1}
\widetilde W(X)\;=\;  \left(\eee^{\beta\|\Phi(X)\|}-1\right)\exp\bigl[|X|(\eee^a-1)\bigr]\;.
\end{equation}
We can now proceed to apply the summability criterion, and bound, given in Proposition \ref{prop.xuan.1} taking into account the exceptional character of $X_0$ whose weight is not given in terms of the Hamiltonian, but
\begin{equation}
\widetilde W(X_0) \;=\; \|F\|\, \exp\bigl[\card{X_0}(\eee^a-1)\bigr]\;.
\end{equation}
The convergence condition \eqref{eq:min.2} requires
\begin{equation}\label{eq:min.2}
\widetilde W(X)\;\le\; \frac{\zeta(X)}{ \displaystyle\prod_{\substack{Y\in\bonds ,\,Y\neq X\\Y\cap X\neq\emptyset}}\bigl[1+\zeta(Y)\bigr]}
\end{equation}
for all $X\in\B_\lat$, for some $\zeta:\bonds \to (0,\infty)$.  This is, precisely, condition \eqref{eq:min.6.exp}.  For $X_0$ the optimal choice is [c.f.\ \eqref{eq:opera.it}]
\begin{equation}
\zeta(X_0)\;=\; \widetilde W(X_0) \,\widetilde T_1(X_0)\;=\; \|F\|\, \exp\bigl[\card{X_0}(\eee^a-1)\bigr] \prod_{\substack{Y\in\bonds \\ Y\cap X_0\neq\emptyset}} \bigl[1+\zeta(Y)\bigr]\;.
\end{equation}
The bound \eqref{eq:min.3} applied to $X_0$ implies then
\begin{equation}\label{eq:min.3.00}
\widetilde T(X_0)\;\le\; \frac{\zeta(X_0)}{\widetilde W(X_0)} \;=\; \prod_{\substack{Y\in\bonds \\ Y\cap X_0\neq\emptyset}} \bigl[1+\zeta(Y)\bigr]
\end{equation}
which, together with \eqref{eq:opera.2} yields \eqref{eq:g.M}.  An alternative way to prove \eqref{eq:min.3.00} is to note that the recursive apporach exploitend in Appendix \ref{app.te} implies that the function 
\begin{equation}
\boldsymbol {\widetilde T}\bigl(\boldsymbol {\widetilde W}\bigr): \B_\lat \longrightarrow [0,\infty)\quad,\quad
\boldsymbol {\widetilde T}\bigl(\boldsymbol {\widetilde W}\bigr)(X)= \widetilde T(X)
\end{equation}
satisfies
\begin{equation}
\boldsymbol {\widetilde T}(\boldsymbol {\widetilde W})(X_0)\;\le\; 
\boldsymbol {\widetilde T}_{\boldsymbol 1}\bigl(\boldsymbol {\widetilde W}\boldsymbol {\widetilde T}\bigr)(X_0)
\;\le\; \boldsymbol {\widetilde T}_{\boldsymbol 1}\bigl(\boldsymbol \zeta\bigr)(X_0)\;=\; 
\prod_{\substack{Y\in\bonds \\ Y\cap X_0\neq\emptyset}} \bigl[1+\zeta(Y)\bigr]\;. \qed
\end{equation}

\section{Approach based on Kirkwood-Salzburg equations}\label{sec:rr.KS-quantum}
In the rest of this section, we show how the approach based on the Kirkwood-Salzburg equations. In this section, we base on the results provided by Bissacot-Fern\'andez-Procacci in \cite{BFP10}, and Park in \cite{Park82}.  In classical mechanics, Gruber and Kunz obtained their analyticity results by setting up the linear equation for the reduced correlations $g_{\Lambda}^\beta(X)$ define by \eqref{eq:conse5}. Such  set of linear equations involves a $\Lambda$-\emph{independent} operator $K$. This linear equation is called the \emph{Kirkwood-Salzburg equations}. In \cite{Park82}, we would like to present the Kirkwood-Salzburg equations type in quantum mechanics.

Let $f$ be a function defined on the set of finite subsets of $\mathbb{Z}^d$. Such functions form a Banach space $\mathcal{F}_{\boldsymbol{\xi}}$:
\begin{equation}\label{eq:KS1}
\mathcal{F}_{\boldsymbol{\xi}}=\left\{f:\|f\|=\sup_{X}\boldsymbol{\xi}^{-X}|f(X)|<\infty\right\},
\end{equation}
where $\bbd{\xi}=\{\xi_x\}_{x\in\Lambda}$ with each $\xi_x>0$ and we abbreviate $\bbd{\xi}^{-X}=\prod_{x\in X}\xi_x^{-1}$. 

We fix a point $x_0\in X$. Then, since 
\begin{eqnarray}
Z\left(\{1\}_{\pro\left(\Lambda\setminus (X\setminus x_0)\right)}\right)&=&\bigg[\prod_{\substack{Y\subset \Lambda\setminus \left(X\setminus x_0\right)\\Y\ni x_0}}(\delta^Y+\epsilon^Y)\bigg]Z\left(\{1\}_{\pro\left(\Lambda\setminus (X\setminus x_0)\right)}\right)\nonumber\\&=&Z\left(\{1\}_{\pro\left(\Lambda\setminus X\right)}\right)+\tilde{Z}\left(\{1\}_{\pro(\Lambda\setminus X)}\right)
\end{eqnarray}
where
\begin{equation}\label{eq:KS2}
\tilde{Z}\left(\{1\}_{\pro(\Lambda\setminus X)}\right):=\sum_{\substack{\{Y_1,\ldots, Y_n\}\\Y_i\ni x_0,\, i=1,\cdots, n}}\prod_{i=1}^n\delta^{Y_i}Z\left(\{1\}_{\pro\left(\Lambda\setminus (X\setminus x_0)\right)}\right)
\end{equation}
it follows that 
\begin{equation}\label{eq:KS3}
Z\left(\{1\}_{\pro(\Lambda\setminus X)}\right)=Z\left(\{1\}_{\pro(\Lambda\setminus (X\setminus x))}\right)-\tilde{Z}\left(\{1\}_{\pro(\Lambda\setminus X)}\right).
\end{equation}
We then have 
\begin{align}
\tilde{Z}\left(\{1\}_{\pro(\Lambda\setminus X)}\right)&\;=\;\sum_{\substack{\{Y_1,\ldots, Y_n\}\\Y_i\ni x_0,\, i=1,\cdots, n}}\prod_{i=1}^n\delta^{Y_i}\prod_{W\subset \Lambda\setminus X}(\delta^{W}+\epsilon^W)Z\left(\{1\}_{\pro\left(\Lambda\setminus (X\setminus x_0)\right)}\right)\nonumber\\&\;=\;\sum_{\substack{\{Y_1,\ldots, Y_n\}\\Y_i\ni x_0,\, i=1,\cdots, n}}\sum_{\{W_1,\ldots, W_m\}}\prod_{i=1}^n\delta^{Y_i}\prod_{j=1}^m\delta^{W_j}Z\left(\{1\}_{\pro\left(\Lambda\setminus (X\setminus x_0)\right)}\right)\nonumber\\\label{eq:KS4}&\;=\;\sum_{\substack{\emptyset\ne S\subset  \Lambda\setminus  (X\setminus x_0)\\S\ni x_0}}K^{\beta}(\{x_0\},S;Z)Z(\{1\}_{\pro(\Lambda\setminus (X\cup S))}).
\end{align}
In order to write  this in the terms of $\Lambda$-independent operator, it is necessary to include the restriction $X\subset \Lambda$ as a factor,
so to extend the functions $g_{\Lambda}$, defined only  when $X\subset \Lambda$ to all $X\subset \mathbb{Z}^d$. Let us define
\begin{equation}\label{eq:KS5}
\chi_{\Lambda}(X)\;=\;\one{X\subset \Lambda}
\end{equation}
and denote 
\begin{equation}\label{eq:KS6}
\tilde{g}_{\Lambda}(X)\;=\;\chi_\Lambda(X)g_{\Lambda}(X).
\end{equation}
From \eqref{eq:KS3} and \eqref{eq:KS4} we obtain
\begin{equation}\label{eq:KS7}
\tilde g_{\Lambda}(X)\;=\;\begin{Bmatrix}
\chi_{\Lambda}(X)\tilde{g}_{\Lambda}(X\setminus x_0)\\\chi_{\Lambda}(X) 

\end{Bmatrix}-\chi_\Lambda(X)\sum_{\substack{\emptyset\ne S\subset\Lambda\setminus (X\setminus x_0)\\S\ni x_0}}K^{\beta}(\{x_0\},S;Z)\tilde{g}_{\Lambda}(X\cup S)
\end{equation}
$K$ is the linear operator on the space of complex-valued functions on $\mathcal{F}_{\bbd{\xi}}$ defined by 
\begin{align}\label{eq:KS8}
(Kf)(\emptyset)&\;=\;0\nonumber\\
(Kf)(X)&\;=\;f(X\setminus x_0)-\chi_{\Lambda}(X)\sum_{\substack{\emptyset\ne S\subset\Lambda\setminus (X\setminus x_0)\\S\ni x_0}}K^{\beta}(\{x_0\},S;Z)f(X\cup S)
\end{align}
Let $\mathbb{1}$ be the element in $\mathcal{F}_{\bbd{\xi}}$ defined by $\mathbb{1}(X)=\one{X=\emptyset}$. In this way we conclude that the function 
\[\tilde g_{\Lambda}:\pro(\mathbb{Z}^d)\longrightarrow \mathbb{C}\]
satisfies the linear condition
\begin{equation}\label{eq:KS9}
\tilde{g}_{\Lambda}=\chi_{\Lambda}\mathbb{1}+\chi_{\Lambda}K\tilde{g}_{\Lambda}
\end{equation}
We have 
\begin{eqnarray}
\left|(Kf)(X)\right|&\le& \bbd{\xi}^{X\setminus x_0}\|f\|_{\bbd{\xi}}+\sum_{\emptyset\ne S\subset \Lambda\setminus (X\setminus x_0)}|K^{\beta}(\{x_0\},S;Z)|\bbd{\xi}^{X\cup S}\|f\|_{\bbd{\xi}}\nonumber\\&\le&\bbd{\xi}^{X}\frac{\|f\|_{\bbd{\xi}}}{\xi_{x_0}}\Bigg[1+\sum_{x_0\in S\in\pro(\Lambda)}\rho_{\B_X}\eee^{a|X|}\Bigg]
\end{eqnarray}
Therefore $K$ is a bounded operator in Banach space $\mathcal{F}_{\bbd{\xi}}$ with the norm bounded by
\begin{equation}\label{eq:KS10}
\|K\|_{\bbd{\xi}}\le \sup_{x_0\in\mathbb{Z}^d}\frac{1}{\xi_x}\Bigg[1+\sup_{x_0}\sum_{x_0\in S\in\pro(\mathbb{Z}^d)}\rho_{\B_X}\eee^{a|X|}\Bigg].
\end{equation}
If $\|K\|_{\bbd{\xi}}<1$, \eqref{eq:KS9} has a unique solution in the Banach space $\mathcal{F}_{\bbd{\xi}}$ given by
\begin{equation}\label{eq:KS11}
\tilde{g}_{\Lambda}=\left[1-\chi_{\Lambda}K\right]^{-1}\chi_{\Lambda}\mathbb{1}.
\end{equation}
By  construction, this solution is analytic in $\beta$ and furthermore 
\begin{equation}\label{eq:KS12}
\|\tilde g_{\Lambda}\|_{\bbd{\xi}}\le \left(1-\|K\|_{\bbd{\xi}}\right)^{-1}.
\end{equation}
As the condition $\|K\|_{\bbd{\xi}}<1$ is independent of $\Lambda$ \eqref{eq:KS11} makes sense in the limit $\Lambda\uparrow\mathbb{Z}^d$ and yields the convergence $\tilde{g}_{\Lambda}\to \tilde{g}$ where the latter is the unique solution of \eqref{eq:KS12} without the the factors $\chi_\Lambda$. Choosing $\xi_{x_0}=\eee^a$ we see that the condition 
\begin{equation}\label{eq:KS13}
\frac{1}{\eee^a}\Bigg[1+\sum_{x_0\in S\in\pro(\Lambda)}\rho_{\B_X}\eee^{a|X|}\Bigg]<1.
\end{equation}
implies the validity of all these properties for $\beta\in \mathcal{D}$ for all $X\in\pro(\mathbb{Z}^d)$, plus analyticity in the interior of $\mathcal{D}$. Using the result in \cite[Proposition 4.1]{BFP10} we can extend the convergence region so to include equality in \eqref{eq:KS13}.
\medskip

\paragraph{\textbf{Acknowledgments}} T. X. Nguyen was partially supported by the grant of GSSI (Gran Sasso Science Institute) during this project. Also, both authors would like to acknowledge the support of the NYU-ECNU Institute of Mathematical Sciences at NYU Shanghai.

\appendix
\section{A gentle introduction to KMS states and Araki-Ion's Gibbs condition}\label{app:gibbs-KMS}

This topic is not easy to access to the uninitiated because available references (e.g.\ \cite{araion74}, \cite[Section III.3]{isr76}, \cite[Sections 5.3--5.4]{brarob2}, \cite[Sections IV.4--IV.5]{Simon93} ) opt for general and elegant expositions that fail to reveal the naturalness of the underlying ideas. 

\subsection{KMS states} 

While classical Gibbs states can be defined through DLR equations (see e.g.\ \cite[Section III.1]{isr76}, \cite[Chapter 2]{geo11}, \cite[Chapter 6]{FV17}), quantum states are defined by their tracial properties.  For inspiration, it is useful to start by the well known finite-volume states.  The tracial characterization of the free ---that is, no interacting--- state is contained in the following equivalence.

\begin{lem} \cite[Lemma IV.4.1]{Simon93}
If $\Theta^0_\Lambda$ is a linear functional on $\alg_\Lambda$, then
 \begin{equation}
\Theta^0(AB)=\Theta_0(BA)\;,\; \Theta^0(\unit)=1\quad \Longleftrightarrow\quad \Theta^0=\tr_\Lambda
\end{equation}
where the right-hand side is the normalized trace defined in \eqref{eq:rr-trace}.
\end{lem}

The presence of the operator $\eee^{-\beta H_\Lambda}$ changes the tracial properties of the Gibbs states $\pi^\beta_\Lambda$ defined in \eqref{eq:rr.qq.2}.  The right property emerges from the following calculation
\begin{equation}
\tr_\Lambda \bigl(A B \,\eee^{-\beta H_\Lambda}\bigr)\;=\;
\tr_\Lambda \bigl(B \,\eee^{-\beta H_\Lambda}\,A\bigr)\;=\;
\tr_\Lambda \bigl(B \,\bigl[\eee^{-\beta H_\Lambda}\,A \,\eee^{\beta H_\Lambda}\,\bigr] \,\eee^{-\beta H_\Lambda}\bigr)\;.
\end{equation}
Therefore,
\begin{equation}\label{eq:app.1}
\pi^\beta_\Lambda(AB)\;=\; \pi^\beta_\Lambda\bigl( B\; \Delta^{\beta\Phi}_\Lambda (A)\bigr)
\end{equation}
with 
\begin{equation}\label{eq:app.0}
\Delta^{\beta\Phi}_\Lambda (A)\;=\; \eee^{-\beta H_\Lambda}\,A \,\eee^{\beta H_\Lambda}\;.
\end{equation}
The passage to infinite volume relies on the following theorem due, with increasing levels of generality, to Streater, Robinson and Ruelle (see references in the paragraph above the reference given for the theorem)

\begin{thm} \cite[Theorem IV.3.3]{Simon93}
If $\Phi\in \mathbb B^{(\alpha)}$ for some $\alpha>0$in\, then there exists a $\star$-automorphism $\Delta^{\beta\Phi}$ on $\alg$ such that 
\begin{equation}
\bigl\| \Delta^{\beta\Phi}_\Lambda(A) -  \Delta^{\beta\Phi}(A)\bigr\|\;\tende{\Lambda\to\lat}\;0
\end{equation}
for each $A\in\alg$.
\end{thm}
The operator $\Delta^{\beta\Phi}$ is defined by an expansion in iterated commutators.
Here is the infinite-volume version of \eqref{eq:app.1}.

\begin{defin}\label{def:KMS}
A state $\Theta$ on $\alg$ is a \empbf{KMS state} for an interaction $\Phi\in\mathbb B^{(\alpha)}$, $\alpha>0$, if
\begin{equation}\label{eq:app.2}
\Theta(AB)\;=\; \Theta\bigl( B\; \Delta^{\beta\Phi} (A)\bigr)
\end{equation}
for each $A,B\in \alg$.
\end{defin}
Given the continuity of states, \eqref{eq:app.2} is equivalent to
\begin{equation}\label{eq:app.3}
\Theta(AB)\;=\; \lim_{\Lambda\to\lat}\Theta\bigl( B\; \Delta^{\beta\Phi}_\Lambda (A)\bigr)\;.
\end{equation}
This expression may be more suitable for concrete arguments and computations, given the explicit character \eqref{eq:app.0} of the finite-volume approximations.  The proof that KMS becomes DLR for classical interactions can be found in \cite[Section III.5]{isr76}.

\subsection{Gibbs condition}

The previous notion of KMS states, even when it fully characterizes what a quantum statistical mechanical state should be, has two drawbacks.  The first one is that does not lead to a direct proof that translation-invariant states coincide with the solutions of a variational approach.  The second one is that, unlike DLR equations, it does not make reference to finite-volume objects and relations (other than that contained in the limit \eqref{eq:app.3}).  An alternative approach, due to Araki and Ion \cite{araion74}, solves both issues.

A natural way to mimic the classical case would be to consider all limits of states $\pi^\beta_\Lambda(\bullet \mid \Theta)$ for all possible quantum boundary conditions $\Theta$.  In the classical case, these limits attains al extremal Gibbs states, and hence they are sufficient to generate the full phase diagram.  No similar result has been proved, however, for extremal KMS states, and the proposed approach is, instead, based on a factorization idea.  The derivation of the DLR equations relies in the decomposition
\begin{equation}
\eee^{-\beta H_{\Lambda'}}\;=\; \eee^{-\beta H_{\Lambda}}\, \eee^{-\beta W_{\Lambda,\Lambda'}} \,\eee^{-\beta H_{\Lambda'\setminus\Lambda}}
\end{equation}
for $\Lambda'\supset\Lambda$.  Here
\begin{equation}\label{eq:app.4}
W_{\Lambda,\Lambda'}\;=\; \sum_{X\in\pro_{\partial\Lambda}\cap\pro_{\Lambda'}}  \Phi(X)\;
\end{equation}
The factorization \eqref{eq:app.4} does not hold in the quantum case because the corresponding operators do not commute.  The following rewriting, however, is valid also for quantum system
\begin{equation}\label{eq:app.5}
\eee^{-\beta (H_{\Lambda'}-W_{\Lambda,\Lambda'})}\;=\; \eee^{-\beta H_{\Lambda}}\, \eee^{-\beta H_{\Lambda'\setminus\Lambda}}
\end{equation}
because operators with disjoint support commute.  To make sense of this factorizationn when $\Lambda'\to \lat$, each exponential must be associated to a homomorphism $\Delta^\beta$ and the resulting decomposition should be interpreted in the following way:  If, for each finite $\Lambda$ we ``perturb" a KMS state $\Theta$ so to ``substract" $W_\Lambda$, the resulting state becomes the product of a $\Delta^{\beta\Phi}_\Lambda$-KMS state ---namely $\pi^\beta_\Lambda$ --- and a KMS state for $\Delta^{\beta\Phi}_{\lat\setminus\Lambda}$.  Denoting $\Theta^{-W_\Lambda}$ the proposed perturbation, \eqref{eq:app.5} transcribes into the identity
\begin{equation}\label{eq:app.6}
\Theta^{-W_\Lambda}\;=\; \pi_\Lambda^\beta \otimes \widetilde\Theta_{\lat\setminus\Lambda}
\end{equation}
with $\widetilde\Theta_{\lat\setminus\Lambda}$ a state on $\alg_{\lat\setminus\Lambda}$.  To define the perturbed state $\Theta^{-W_\Lambda}$, let us again look first on finite volumes.  In this case, to substract $W_{\Lambda,\Lambda'}$ in the state $\pi^\beta_{\Lambda'}$ means to pass to
\begin{equation}
\pi^{-W_{\Lambda,\Lambda'}}_{\Lambda'}(A)\;=\;  \frac{\tr_{\Lambda'} \bigl(A \,\eee^{-\beta H_\Lambda}\,\eee^{-\beta H_{\Lambda'\setminus\Lambda}}\bigr)}{\tr_{\Lambda'} \bigl(\eee^{-\beta H_\Lambda}\,\eee^{-\beta H_{\Lambda'\setminus\Lambda}}\bigr)} \;=\; \bigl[\pi_\Lambda^\beta\otimes \pi^\beta_{\Lambda'\setminus\Lambda}\bigr](A)\;.
\end{equation}
Such a state can be constructed from $\pi_{\Lambda'}^\beta$ by 
 changing the exponential weight according to the trivial identity
\begin{equation}
\eee^{-\beta (H_{\Lambda'}-W_{\Lambda,\Lambda'})}\;=\; \bigl[\eee^{-\beta (H_{\Lambda'}-W_{\Lambda,\Lambda'})}\; \eee^{\beta H_{\Lambda'}}\bigr]\,\eee^{-\beta H_{\Lambda'}}\;=:\; \Gamma_\Lambda^{\beta,\Lambda'} \,\eee^{-\beta H_{\Lambda'}}\;.
\end{equation}
Therefore, the removal of the term $W_{\Lambda,\Lambda'}$ yields a perturbed state  
\begin{equation}\label{eq:app.10}
\pi^{-W_{\Lambda,\Lambda'}}_{\Lambda'}(A)\;=\; \frac{\pi^\beta_{\Lambda'}\bigl(A\,\Gamma_\Lambda^{\beta,\Lambda'} \bigr)}{\pi^\beta_{\Lambda'}\bigl(\Gamma_\Lambda^{\beta,\Lambda'}\bigr) }\;.
\end{equation}
A time-ordered expansion (see, e.g., the proof of Theorem IV.5.5 in \cite{Simon93}) shows that the operators $\Gamma_\Lambda^{\beta,\Lambda'}$ converge, as $\Lambda' \to \lat$, to a bounded operator 
\begin{equation}
\Gamma^\beta_\Lambda\;=\; \lim_{\Lambda'\to\lat} \eee^{-\beta (H_{\Lambda'}-W_{\Lambda,\Lambda'})}\; \eee^{\beta H_{\Lambda'}}
\end{equation}
which is used to define the infinite-volume analogue of \eqref{eq:app.10}

\begin{defin}
The $-W_\Lambda$-\empbf{perturbation} of a state $\Theta$ is the state such for each $A\in\alg$
\begin{equation}\label{eq:app.11}
\Theta^{-W_\Lambda}(A)\;=\; \frac{\Theta\bigl(A\,\Gamma^\beta_\Lambda\bigr)}{\Theta\bigl(\Gamma^\beta_\Lambda\bigr)}
\;=\;  \lim_{\Lambda'\to\lat} \frac{\Theta\bigl(A\,\eee^{-\beta (H_{\Lambda'}-W_{\Lambda,\Lambda'})}\; \eee^{\beta H_{\Lambda'}}\bigr)}{\Theta\bigl(\eee^{-\beta (H_{\Lambda'}-W_{\Lambda,\Lambda'}\bigr)}\; \eee^{\beta H_{\Lambda'}}\bigr)}
\end{equation}
\end{defin}
It is relatively simple to verify that if $\theta$ is a $\Delta^{\beta\Phi}$-KMS state, then $\Theta^{-W_\Lambda}$ is a
$\bigl(\Delta^{\beta\Phi}_\Lambda\,\Delta^{\beta\Phi}_{\lat\setminus\Lambda}\bigr)$-KMS state.

\begin{defin}
A state $\Theta$ on $\alg$ is called a \empbf{Gibbs state} for an interaction $\Phi\in\mathbb B^{(\alpha)}$, for some $\alpha>0$, if for each finite $\Lambda$ there exists a state $\widetilde\Theta_{\lat\setminus\Lambda}$ on $\alg_{\lat\setminus\Lambda}$ such that \eqref{eq:app.6} holds.
\end{defin}

A more constructive version of this definition is obtained by ``perturbing back'' the factorized state into the original $\Theta$ through the inverse of the first identity in \eqref{eq:app.11}:
\begin{equation}\label{eq:app.12}
\Theta(A)\;=\; \frac{\Theta^{-W_\Lambda}\bigl(A\,\bigl[\Gamma^\beta_\Lambda\bigr]^{-1}\bigr)}{\Theta^{-W_\Lambda}\bigl(\bigl[\Gamma^\beta_\Lambda\bigr]^{-1}\bigr)}
\end{equation}
with
\begin{equation}
\bigl[\Gamma^\beta_\Lambda\bigr]^{-1}\;=\; \lim_{\Lambda'\to\lat} \eee^{-\beta H_{\Lambda'}}\,\eee^{\beta (H_{\Lambda'}-W_{\Lambda,\Lambda'})}\; 
\end{equation}
In this way, we obtain the following, more cumbersome but constructive definition (which is actually the original definition given by Araki and Ion in their seminal work \cite{araion74}).

\begin{prop}[Alternative definition of Gibbs states]
A state $\Theta$ on $\alg$ is a \empbf{Gibbs state} for an interaction $\Phi\in\mathbb B^{(\alpha)}$, for some $\alpha>0$, if, and only if, for each finite $\Lambda$ there exists a state $\widetilde\Theta_{\lat\setminus\Lambda}$ on $\alg_{\lat\setminus\Lambda}$ such that
\begin{eqnarray}
\Theta(A) &=& \frac{\pi_\Lambda^\beta \otimes \widetilde\Theta_{\lat\setminus\Lambda}\bigl(A\,\bigl[\Gamma^\beta_\Lambda\bigr]^{-1}\bigr)}
{\pi_\Lambda^\beta \otimes \widetilde\Theta_{\lat\setminus\Lambda}\bigl(\bigl[\Gamma^\beta_\Lambda\bigr]^{-1}\bigr)}\\
&=& \lim_{\Lambda'\to\lat} \frac{\pi_\Lambda^\beta \otimes \widetilde\Theta_{\lat\setminus\Lambda}\bigl(A\,\eee^{-\beta H_{\Lambda'}}\,\eee^{\beta (H_{\Lambda'}-W_{\Lambda,\Lambda'})}\bigr)}
{\pi_\Lambda^\beta \otimes \widetilde\Theta_{\lat\setminus\Lambda}\bigl(\eee^{-\beta H_{\Lambda'}}\,\eee^{\beta (H_{\Lambda'}-W_{\Lambda,\Lambda'})}\bigr)}
\end{eqnarray}
\end{prop}
The second equality is the working definition adopted in \ref{sec:gibbs-KMS} above.
The proof that the sets of KMS and Gibbs states coincide can be found, for instance, in \cite{araion74}, \cite[Sections 5.4]{brarob2} and \cite[Sections IV.5]{Simon93}.

\section{Disguises of the M\"obius transform}\label{app:decoupling}

The M\"obius transform is often surreptitiously introduced through very elegant alternative treatments.  Let us spell out the two more popular ones.  

\subsection{Park's $\delta$'s and $\varepsilon$'s}

Park \cite{Park82} rewrites the RHS of \eqref{eq:ms.moeb} in terms of operators $\delta^x$ which amounts to a discrete derivative.  To prove the inverse formula in the LHS of this expression, he introduces a sort of inverse operators $\varepsilon^x$.  Here is a transcription of his approach (our $\varepsilon$ is slightly different from Park's).  
\begin{prop}
Consider the following operators on the vector space of functions $F: \{\mbox{parts of } \mathcal S\} \to \mathbb V$ where $\mathcal S$ is a finite set and $\mathbb V$ a vector space.  Define the following operators for each $x\in\Lambda$:  
\begin{eqnarray}
\delta^x F(A) &=& F\bigl(A\cup\{x\}\bigr) - F(A)\\
\varepsilon^x F(A) &=& F\bigl(A\cup\{x\}\bigr) + F(A)
\end{eqnarray}
The following properties are easily verified:
\begin{itemize}
\item[(a)] These operators commute between themselves.
\item[(b)] $\displaystyle \bigl(1+\delta^x\bigr)F(A) = F\bigl(A\cup\{x\}\bigr) =  \bigl(\varepsilon^x-1\bigr)F(A)$ 
\end{itemize}
\end{prop}
[As a matter of fact, Park;s definition of $\delta^x F(A)$ is $F\bigl(A\cup\{x\}\bigr) - F\bigl(A\setminus\{x\}\bigr)$.  Hence, it coincide with ours each time that $x\not\in A$., as in the application below.] 

The commutativity properties justifies the definition of $\delta^S:=\prod_{x\in S} \delta^x$ and $\varepsilon^S:=\prod_{x\in S} \varepsilon^x$, for each $S\subset\mathcal S$.  In particular, part (b) of the proposition implies
\begin{equation}\label{eq:app.c.1}
\bigl(1+\delta\bigr)^SF(\emptyset) = F(S) =  \bigl(\varepsilon-1\bigr)^SF(\emptyset)\;.
\end{equation}

The interest of these operators stems from the following lemma who follows from a two-line induction argument.

\begin{lem}
For each function $F$ and sets $B\subset\mathcal S$
\begin{eqnarray}\label{eq:app.c.1.1}
\delta^BF(\emptyset) &=& \sum_{A\subset B} (-1)^{\card{B\setminus A}} F(A)\\
\varepsilon^BF(\emptyset) &=& \sum_{A\subset B} F(A)\;. \label{eq:app.c.1.2}
\end{eqnarray}
\end{lem}

Identities \eqref{eq:app.c.1}--\eqref{eq:app.c.1.2} readily imply the inclusion-exclusion relation.

\begin{thm}[M\"obius transform]
For each set $S\subset\Lambda$,
\begin{itemize}
\item[(M1)] $\displaystyle G(S)=\delta^S F(\emptyset)\;\Longrightarrow\; \varepsilon^S G(\emptyset)=F(S)$\;.
\item[(M2)] $\displaystyle F(S)=\varepsilon^S G(\emptyset)\;\Longrightarrow\; \delta^S F(\emptyset)=G(S)$\;.
\end{itemize}
\proof

\noindent
\emph{[(M1)]}
\[
\varepsilon^S G(\emptyset)\;=\; \sum_{A\subset S} G(A)\;=\; \sum_{A\subset S} \delta^A F(\emptyset)
\;=\; \bigl(1+\delta\bigr)^S F(\emptyset)\;=\;F(S)\;.
\]

\noindent
\emph{[(M2)]}
\[
\delta^S F(\emptyset)\;=\; \sum_{A\subset S} (-1)^{\card{S\setminus A}} F(A)\;=\; \sum_{A\subset S} (-1)^{\card{S\setminus A}} \varepsilon^A G(\emptyset)
\;=\; \bigl(\varepsilon -1\bigr)^S G(\emptyset)\;=\;G(S)\;.\quad\qed
\]

\end{thm}

\subsection{Use of decoupling parameters}\label{ssec:decoup}

This is a time-honored technique to represent the RHS of \eqref{eq:ms.moeb}.  In our general setup it involves the introduction of a parameter $s_x\in[0,1]$ associated to each point of $\mathcal S$.  A point $x$ is thought as decoupled if $s_x=0$ and fully coupled if $s_x=1$.  The association is made in reference to a given function $F$ on subsets and it is defined so to establish associated function $\widehat F:[0,1]^{\mathcal S} \longrightarrow \mathbb V$ such that
\begin{itemize}
\item[(i)] The function $\underline s \to \widehat F(\underline s)$ is smooth (at least with first-order partial derivatives).
\item[(ii)] For each $A\subset\mathcal S$, $F(A) \;=\; \widehat F\bigl(\unit_A 0_{{\mathcal S}\setminus A}\bigr)$.
\end{itemize}

The example of interest in Park's work is for ${\mathcal S}= \B_\Lambda$ and a function $\widehat F(\B)=Z_{\B}^\Lambda$. The ``partially decoupled" associated function is $\widehat F(\underline s)=Z_{\Lambda}^{s\phi}$ in which is term $\Phi(X)$ is multiplied by a factor $s_X$ that interpolates between decoupling and full coupling.  

These parameters allow to express the relation $G(S)=\delta^S F$ as an application of the fundamental theorem of calculus. 

\begin{prop}
\begin{equation}
\sum_{A\subset B} (-1)^{\card{B\setminus A}} \widehat F\bigl(\unit_{B} 0_{{\mathcal S}\setminus A}\bigr) \;=\;
\Bigl\{\prod_{x\in B}\int_0^1 ds_x \frac{\partial}{\partial s_x}\Bigr\} \widehat F(\underline 0)\;.
\end{equation}
\end{prop}
\proof

By the FTC,
\begin{equation}
\int_0^1 ds_x \frac{\partial}{\partial s_x} \widehat F(\underline s)\;=\; \widehat F\bigl(1_x \, s_{{\mathcal S}\setminus\{x\}}\bigr) - F\bigl(0_x s_{{\mathcal S}\setminus\{x\}}\bigr)\;.
\end{equation}
This operation is, therefore, equivalent to the operator $\delta^x$.  The proof then follows easily by induction on $\card B$.\;\qed


\section{Summability of tree expansions} \label{app.te}

The starting step is to fix an order of the vertices $X_i$ of $\mathbb G$.
\begin{description}
\item[\bf Generation 0] Choose $X_0$ as root and consider the resulting graph distance (= minimal number of links needed to attain the root) of the remaining vertices $X_i$.   Let $\mathcal K_k$ denote the set of vertices at graph distance $k$ from the root.  The tree $\mathcal T$ keeps the vertices at their respective distances, it only removes some of the links in $\mathbb G$.
\item[\bf Generation 1] Label the different vertices in $\mathcal K_1$ in the form $X_{(1,i_1)}$ and keep the corresponding edges $(X_0,X_{(1,i)})$.  The labels $(1,i)$ are chosen in increasing vertex order.  All edges in $\mathbb G$ linking siblings, that is bonds of the form $(X_{(1,i_1)},X_{(1,j)})$--- are omitted. 
\item[\bf Generation $\boldsymbol k$ $\boldsymbol{ (2\le k\le n)}$] After determining the tree up to the $(k-1)$-th generation, the vertices in $\mathcal K_{k-1}$ are labelled in the form $X_{(1,i_1,\ldots,i_{k-1})}$.  We start by the smallest such vertex --which is labelled by the sequence $\boldsymbol i_{k-1}:=(1,i_1,\ldots,i_{k-1})$ which is smallest in lexicographic order--- and keep all the links between it and vertices in $\mathcal K_k$.  Label then in the form $(\boldsymbol i_{k-1}:,i_k)$ with $i_k$ respecting vertex order.  Then continue with the second vertex in $\mathcal K_{k-1}$, including all its links with the remaining vertices in $\mathcal K_k$.  Proceed in this fashion through the successive vertices in $\mathcal K_{k-1}$ following the order of increasing subscripts.  Omit all the remaining edges in $\mathbb G$ linking vertices in $\mathcal K_k$ with vertices in $\mathcal K_{k-1}$ or between vertices in $\mathcal K_k$.  
\end{description}

We see that the tree so obtained are uniquely characterized by the absence of links between (i) vertices at the same graph distance of the root $X_0$, and (ii) a vertex labelled  $\boldsymbol i_{k-1}$ with a vertex labelled $(\boldsymbol i'_{k-1},i_k)$
with $\boldsymbol i_{k-1}<\boldsymbol i'_{k-1}$.  Let us denote $\mathbb T_n^P$ the set of planar labelled trees satisfying both these conditions (``P" stands for \emph{Penrose}, who was the first to propose this construction ---see Appendix \ref{app.te}).  

\begin{prop}[Sum of tree expansions] \label{prop:cast}

\end{prop}

\begin{equation}
T(X_0)\;\le\; 1+ \sum_{n\ge 1} \sum_{\tau\in \mathbb T_n} \sum_{\substack{\{X_1,\ldots, X_n\}\\X_i\in\bonds , X_i\neq X_0}} 
\prod_{i=0}^n c_{s_i}\bigl(X_i,X_{(i,1)},\ldots, X_{(i,s_i)}\bigr)\prod_{j=1}^n W(X_j)
\end{equation}
with
\begin{equation}
c_{s_i}\bigl(X_i,X_{(i,1)},\ldots, X_{(i,s_i)}\;=\; \prod_{j=1}^n\one{X_{(i,j)}\cap X_i \neq\emptyset\,,\,X_{(i,j)}\neq X_i}
\prod_{1\le k<\ell\le n}\one{X_{(i,k)}\neq X_{(i,\ell)}}\;.
\end{equation}  

 \section{General convergence criterion}\label{app:sketch}
 
In fact, the criteria detailed in Section \ref{ssec:rr.tan.10} are a consequence of a more general criterion~\cite{PF07} reviewed in \cite{BFP10,Jan18,FN19}.  We state it here for completeness.

\begin{thm}\label{theo:rr.tan.06.1}
Consider the function $\boldsymbol{\varphi}$ from $[0,+\infty)^{\mathcal{P}}$ to $[0,+\infty]^{\mathcal{P}}$ defined by
\begin{equation}\label{eq:quan26}
\varphi_{Y_0}(\boldsymbol{\mu})\;=\;
1+\sum_{n\ge1}\sum_{\{Y_1,\ldots,Y_n\}\subset\pro}\prod_{i=1}^n\one{Y_0\nsim Y_i}\prod_{1\le k< \ell\le n}\one{Y_k\sim Y_\ell}\prod_{j=1}^n\mu_{Y_j}\;.
\end{equation}
If for a given $\boldsymbol{\lambda}\in[0,+\infty)^{\mathcal{P}}$ there exists $\boldsymbol{\mu}\in[0,+\infty)^{\mathcal{P}}$ such that 
\begin{equation}\label{eq:quan23}
\lambda_{Y}\varphi_{Y}(\boldsymbol{\mu})\le \mu_{Y}
\end{equation}
for each $Y\in\mathcal{P}$, the following holds.
\begin{itemize}
\item[(a)] The expansions \eqref{eq:polmect10.1} converge absolutely and uniformly in $\Lambda$ in the polydisc $\card{\rho_Y}\le \lambda_Y$, $Y\in\pro$.

\item[(b)] For each $\boldsymbol{\lambda}\in[0,\infty)^\pro$ let $\boldsymbol{T}_{\boldsymbol{\lambda}}(\boldsymbol{\mu})=\bbd{\lambda}\bbd{\varphi}(\boldsymbol{\mu})$ be the map from $[0,+\infty)^{\mathcal{P}}$ to $[0,+\infty]^{\mathcal{P}}$ defined by 
\begin{equation}
\bigl[\boldsymbol{T}_{\boldsymbol{\lambda}}(\boldsymbol{\mu})\bigr]_Y\;=\;\lambda_Y\,\bbd{\varphi}_Y(\boldsymbol{\mu})
\end{equation}
for each $Y\in\pro$. Then
\begin{itemize}
\item[(i)] There exist $\bbd{\lambda}^*, \,\bbd{T}_{\bbd{\lambda}}^{\infty}(\bbd{\mu})\in[0,+\infty)^{\mathcal{P}}$ such that 
\begin{equation}\label{eq:quan24}
\bbd{T}^n_{\bbd{\lambda}}(\bbd{\lambda})\underset{n\to\infty}{\nearrow}\bbd{\lambda}^*,\,\bbd{T}^n_{\bbd{\lambda}}(\bbd{\lambda})\underset{n\to\infty}{\searrow}\bbd{T}_{\bbd{\lambda}}^{\infty}(\bbd{\mu})
\end{equation}
and $\bbd{T}_{\bbd{\lambda}}(\bbd{\lambda}^*)=\bbd{\lambda}^*$.

\item[(ii)] 
$\card{\bbd{\Sigma}}=\lim_\Lambda \card{\bbd{\Sigma}}^\Lambda$ exists and satisfies,
for each $n\in\mathbb{N}$,
\begin{equation}\label{eq:polmect10.1.3}
\bbd{\lambda}\card{\bbd{\Sigma}}\!(\bbd{\lambda})\;\le\; \bbd{\lambda}^*\;\le\; \bbd{T}^{\infty}_{\bbd{\lambda}}(\bbd{\mu})\;\le\; \bbd{T}^{n+1}_{\bbd{\lambda}}(\bbd{\mu})\le \bbd{T}^{n}_{\bbd{\lambda}}(\bbd{\mu})\;\le\; \bbd{\mu}.
\end{equation}
\end{itemize}
\end{itemize}
\end{thm}

\section{Comparison with Park's results}\label{sec:appendix}

In this subsection, we compare our estimations with the results provided by Y. M. Park in \cite{Park82}.  To describe the latter, let us denote
\begin{eqnarray}\label{eq:com2}
b_{\mathrm{p}}&:=&\sup_{x\in\mathbb{Z}^d}\sum_{X\ni x}\|\Phi(X)\|\ge \|\Phi(X)\|_{\infty}\\
c_{\mathrm{p}}&:=&\sup_{x\in\mathbb{Z}^d}\sum_{\substack{x\in X\subset \mathbb{Z}^{d}\\X\,\mathrm{finite}}}\|\Phi(X)\|\eee^{(\alpha/2)\,|X|}\\
\|\Phi(X)\|_{\infty}&:=&\sup_{n}\sup_{\substack{x\in X\subset \mathbb{Z}^d\\|X|=n}}\|\Phi(X)\|\\\|\Phi(X)\|_{\alpha/2}&:=&\sup_n\sup_{x\in\mathbb{Z}^d}\sum_{\substack{x\in X\subset \mathbb{Z}^d\\|X|=n}}\|\Phi(X)\|\eee^{\alpha/2|X|}
\end{eqnarray}
%
Park uses Kirkwood-Salzburg equations, rather than full-fledge cluster expansions; as a consequence, he does not address analyticity of free energies but of correlation functions.  
Park's domain of $\beta$-analyticity takes the form
\[\mathcal{D}_{\mathrm{p}}:=\left\{\beta\in\mathbb{C}:|\beta|< \frac{\alpha}{4},\;A(|\beta|)\eee^{-\frac{\alpha}{4}+2b|\beta|}<1\right\}\]
where
\[A(|\beta|)=1+c_1|\beta|\frac{\eee^{\frac{-\alpha}{4}+b_{\mathrm{p}}|\beta|}}{1-\eee^{\frac{-\alpha}{4}+b_{\mathrm{p}}|\beta|}}\;.\]

Let us denote
\[F(\beta)\,:= \,A(\beta)\eee^{-\alpha/4+2\,b_{\mathrm{p}}|\beta|}-1\]
Elementary analysis show that $F(\beta)$ is strictly increasing function, so $F$ has a unique root $\beta^*\in(0,\alpha/(b_{\mathrm{p}}8))$.  As a consequence,
\begin{eqnarray}\label{eq:com4}
c_1|\beta^*|\eee^{b_{\mathrm{p}}|\beta^*|}&=&\left[\eee^{\alpha/4-2\,b_{\mathrm{p}}|\beta^*|}-1\right]\left[\eee^{\alpha/4}-\eee^{b_{\mathrm{p}}|\beta^*|}\right]\;.
\end{eqnarray}

\begin{figure}[htb!]
  \includegraphics[width=\linewidth]{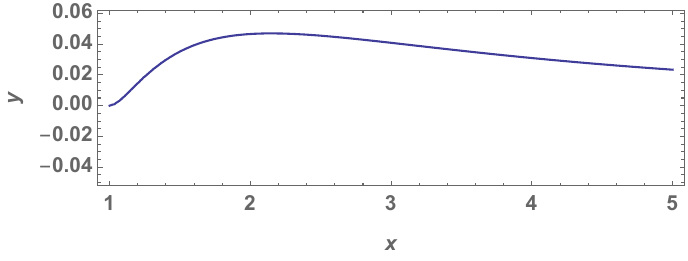}
  \caption{The graphic of the solution $2d\beta$}
  \label{fig:sol1}
\end{figure}

For the sake of simplicity, we focus on lattice systems having only nearest-neighbor pair-interactions. Without loss of generality we assume that $\|\Phi(X)\|_{\infty}=1$, so $b_{\mathrm{p}}=2d$ and $c_1>c_{\mathrm{p}}=2d\eee^{\alpha}$. Replacing $c_1$ by $c_{\mathrm{p}}$ we obtain that the solution of the equation
\begin{eqnarray}\label{eq:addcom5}
2d\eee^{\alpha}|\beta^*|\eee^{2d|\beta^*|}&=&\left[\eee^{\alpha/4-2d|\beta|^*}-1\right]\left[\eee^{\alpha/4}-\eee^{2d|\beta|}\right]
\end{eqnarray}
yields a lower bound on the radius of convergence. 
Figure \eqref{fig:sol1} shows the dependence of $y=2d|\beta|$ on $x=\eee^{\alpha/4}$. 
Note that \eqref{eq:addcom5} implies that $\beta\to 0$ when $|\alpha|\to +\infty$. From the plot we see  that $2d\beta^*<0.06$. Hence,
\[|\Phi\|_{\infty}|\beta|\eee^{\|\Phi\|_{\infty}|\beta|}\approx \frac{0.06}{2d}\left(1+\frac{0.06}{2d}\right)<\frac{0.0921428}{2d}.\]
The last result is our bound obtained in previous subsection.

\end{document}